\documentclass[prx,twocolumn,superscriptaddress,nofootinbib,longbibliography]{revtex4-2}
\usepackage{graphicx}% Include figure files
\usepackage{bm}% bold math
\usepackage{amssymb}
\usepackage{color}
\usepackage{amsmath}
\usepackage{mathrsfs}
\usepackage{amstext}
\usepackage[english]{babel}
\usepackage{latexsym}
\usepackage{array}             % Tabular
\usepackage{multirow}         % Multicolumn Tables
\usepackage[usenames,dvipsnames]{xcolor}
\usepackage[colorlinks=true,citecolor=Blue,linkcolor=RubineRed,urlcolor=Blue]{hyperref}
\usepackage{physics}
\usepackage{enumerate}

\newcommand{\pap}[1]{\left( #1 \right)}
\newcommand{\pas}[1]{\left[#1 \right]}

\begin{document}
\title{R\'enyi Entropy Singularities as Signatures of Topological Criticality in Coupled Photon-Fermion Systems}
\author{F. P. M. M\'endez-C\'ordoba\,\href{https://orcid.org/0000-0002-7685-2264}{\includegraphics[scale=0.45]{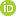}}}
\email{fp.mendez10@uniandes.edu.co}
\affiliation{Departamento de F\'isica, Universidad de Los Andes, A.A. 4976, Bogot\'a, Colombia}

\author{J. J. Mendoza-Arenas\,\href{https://orcid.org/0000-0002-2214-334X}{\includegraphics[scale=0.45]{orcid}}}
\affiliation{Departamento de F\'isica, Universidad de Los Andes, A.A. 4976, Bogot\'a, Colombia}

\author{\\F. J. G\'omez-Ruiz\,\href{https://orcid.org/0000-0002-1855-0671}{\includegraphics[scale=0.45]{orcid}}}
\affiliation{Donostia International Physics Center,  E-20018 San Sebasti\'an, Spain}
\affiliation{Departamento de F\'isica, Universidad de Los Andes, A.A. 4976, Bogot\'a, Colombia}

\author{F. J. Rodr\'iguez\,\href{https://orcid.org/0000-0001-5383-4218}{\includegraphics[scale=0.45]{orcid}}}
\affiliation{Departamento de F\'isica, Universidad de Los Andes, A.A. 4976, Bogot\'a, Colombia}

\author{C. Tejedor\,\href{https://orcid.org/0000-0002-4577-358X}{\includegraphics[scale=0.45]{orcid}}}
\affiliation{Departamento de F{\'i}sica Te{\'o}rica de la Materia Condensada and Condensed Matter Physics Center (IFIMAC), Universidad Aut{\'o}noma de Madrid, Madrid 28049, Spain}

\author{L. Quiroga\,\href{https://orcid.org/0000-0003-2235-3344}{\includegraphics[scale=0.45]{orcid}}}
\affiliation{Departamento de F\'isica, Universidad de Los Andes, A.A. 4976, Bogot\'a, Colombia}

\begin{abstract}
\begin{center}
(Received 20 July 2020; accepted 13 October 2020; published 20 November 2020)
\end{center}
We show that the topological phase transition for a Kitaev chain embedded in a cavity can be identified
by measuring experimentally accessible photon observables such as the Fano factor and the cavity quadrature amplitudes. Moreover, based on density matrix renormalization group numerical calculations, endorsed by an analytical Gaussian approximation for the cavity state, we propose a direct link between those observables and quantum entropy singularities. We study two bipartite entanglement measures, the von Neumann and R\'enyi entanglement entropies, between light and matter subsystems. Even though both display singularities at the topological phase transition points, remarkably only the R\'enyi entropy can be analytically connected to the measurable Fano factor. Consequently, we show a method to recover the bipartite entanglement of the system from a cavity observable. Thus, we put forward a path to experimentally access the control and detection of a topological quantum phase transition via the Rényi entropy, which can be measured by standard low noise linear amplification techniques in superconducting circuits. In this way, the main quantum information features of Majorana polaritons in photon-fermion systems can be addressed in feasible experimental setups.\\
\\
DOI: \href{https://journals.aps.org/prresearch/abstract/10.1103/PhysRevResearch.2.043264}{10.1103/PhysRevResearch.2.043264}
\end{abstract}
\maketitle
\section{Introduction}
The understanding of correlated matter strongly coupled to quantum light has been an intense area of research both theoretically and experimentally in the last few years. Hybrid photonic technologies for control of complex systems have been constantly improving, now acting as cornerstones for quantum simulations in cutting-edge platforms such as optical lattices. Namely, trapped ions are subjected to high control by laser beams allowing the manipulation of the main system parameters~\cite{ExtendedBH, Lewis_SwanDM, REHubbard&Spin,BrydgesRERandomized,Camacho_Guardian_2017OpticalLattices}. Strong light-matter couplings have been generated in superfluid and Bose-Einstein gases embedded in cavities now available to study systems with exquisitely tailored properties~\cite{ExpFERMBOSONS,ExpDickeSuperfluid,ExpDickeSuperfluidPRL,L_onard_2017BECCAvity,roux2020nat}. Furthermore, the analysis of light-controlled condensed matter systems has led to predictions of a rich variety of phenomena, including the enhancement of electron-photon superconductivity by cavity mediated fields~\cite{kiffner2019prb,thomasEFSC,curtis2019prl,FrankEPSC,GaoPEP}. Experimentally, new physical features as well as control opportunities in the ultrastrong and deep-strong-coupling regimes, where coupling strengths are comparable to or larger than subsystem energies, have been observed recently using circuit quantum electrodynamics microwave cavities~\cite{FornRMP2019,FriskNPR2019}.\\
\\
Motivated by these remarkable advances, we are encouraged to establishing new feasible hybrid cavity scenarios for the detection and control of nonlocal correlated features in solid-state setups such as topological materials~\cite{DartiailhMajoranaPairs,frank2019prl,Nie_2020Superatom}. A great deal of attention has been recently devoted to assessing nonlocal Majorana fermion quasiparticles in chains with strong spin-orbit coupling disposed over an {\it s}-wave superconductor~\cite{Mourik1003,ExpMajorana,Exp2016,Exp2018}. Majorana fermions, as topological quasiparticles in solid-state environments, have been widely searched due to their unconventional properties against local decoherence and hence for possible technological solutions to fault-tolerant quantum computing protocols~\cite{FernandoTwoTimeCorrelations,DynamicalDelocalizationMajorana,EntanglementInManyBody,AasenMajoranaQuantumComputing}.\\
\\
Since the seminal work by Kitaev~\cite{Kitaev} where a one-dimensional spinless fermion chain was shown to feature Majorana physics, topological properties of hybrid semiconductor-superconductor systems~\cite{Mourik1003,ExpMajorana,Exp2016,Exp2018} have been explored looking for the presence of the so-called zero energy modes  (ZEM), corresponding to quasiparticles localized at the boundaries of the chain. The fact that these quasiparticles have zero energy makes them potential candidates for the implementation of non-Abelian gate operations within two dimensional (2D) arrangements~\cite{majoranareturns,majorana,AnyonsBurton,SurvivalMajorana,ProgramableMajoranas}. 
However, some open questions still remain about the experimental occurrence of these modes since the reported phenomena observed in those experiments could be caused by a variety of alternative competing effects~\cite{andreevKim}. Therefore, new experimental frames are highly desirable to find unambiguous signs of such quasiparticles.\\
\\
An important question in this context is whether the topological phase transition of Majorana polaritons, for instance in a fermion chain embedded in a microwave cavity~\cite{MirceaTrifHamiltonian,trif2019prl,TrifMajoranasSpinOrbit}, can be detected by accessing observables such as the mean number of photons, field quadratures or cavity Fano factor (FF). In this paper, we report on an information-theoretic approach based on the analysis of the R\'enyi entropy ($S_{\rm R}$) of order two between light and matter subsystems, for connecting its singular behavior, resulting from the topological transition, with the FF.  Consequently, we show a path to characterize the bipartite entanglement of the light-matter system and how to use it as a witness to identify quantum phase transitions. Additionally, we show that in a wide parameter coupling regime the cavity state is faithfully represented by a Gaussian-state (GS). Within this description, measurements of the Fano parameter and single-mode quadrature amplitudes yield directly to assessing the R\'enyi entropy. This approach allows us to link directly accessible microwave observables to quantum light-matter correlations~\cite{Acevedo2_2015,Acevedo_2015,Gomez2018}, and clarifies the role of topological phases hosted by cavity-fermion coupled systems.\\
\\
Our paper is organized as follows. Section~\ref{Sec1} gives the description of the Kitaev model embedded in a microwave cavity. In Sec.~\ref{Sec:MF}, we present the mean-field approach of the system, which is useful to predict the response of the cavity. In Sec.~\ref{Sec:PhaseDiagram}, we present the phase diagram of the composite system obtained numerically. In Sec.\ref{Sec:VNCRiticalityGS} we show that the composite system signals the phase transitions in the von Neumann entropy and that the state of the cavity can be approximated by a single-mode Gaussian-state. In Sec. \ref{Sec:Renyi} we show the connection between the Fano factor and the Rényi entropy.
Finally, in Sec.\ref{Sec:Conclusions} we present a summary of our work.

\begin{figure}[t!]
\begin{center}
\includegraphics[width=1\linewidth]{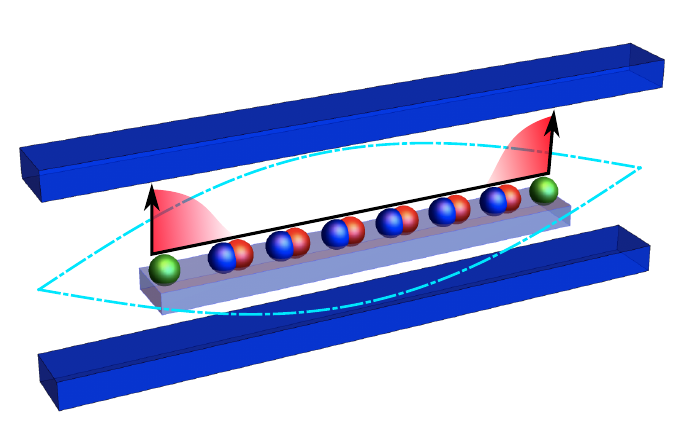}
\caption {Schematic illustration of a Kitaev chain embedded in a single-microwave cavity. The blue curve denotes the profile of the fundamental mode of the cavity. Majorana fermion quasiparticles are depicted as blue and red spheres (bulk) and as green spheres for the edge unpaired quasiparticles for an isolated Kitaev chain in the topological phase. The light red regions illustrate the hybridization effect yielding to Majorana polaritons.
}\label{fig:Sketch}
\end{center}
\end{figure}

\section{Photon-Fermion Model}\label{Sec1}
We consider a Kitaev chain embedded in a single-mode microwave cavity as schematically shown in Fig. \ref{fig:Sketch}. The system is described by the Hamiltonian 
\begin{equation}\label{FullHamiltonian}
\hat{\mathcal{H}}=\hat{\mathcal{H}}_{\rm C}+\hat{\mathcal{H}}_{\rm K}+\hat{\mathcal{H}}_{{\rm Int}}.
\end{equation}
Here $\hat{\mathcal{H}}_{\rm C}=\omega \hat{a}^\dagger \hat{a}$ is the Hamiltonian describing the microwave single-mode cavity, with $\hat{a} \pap{\hat{a}^{\dagger}}$ the annihilation (creation) microwave photon operator, and $\omega$ is the energy of the cavity; we set the energy scale by taking $\omega=1$. The isolated open-end Kitaev chain Hamiltonian $\hat{\mathcal{H}}_{\rm K}$ is given by

\begin{equation}
\begin{split}
\hat{\mathcal{H}}_{\rm K}=&-\frac{\mu}{2} \sum_{j=1}^L\left[2\hat{c}_j^\dagger \hat{c}_j-\hat{1}\right] -t\sum_{j=1}^{L-1}\left[\hat{c}_{j}^{\dagger}\hat{c}_{j+1}+\hat{c}_{j+1}^{\dagger}\hat{c}_{j}\right]\\   &  +\Delta \sum_{j=1}^{L-1}\left[\hat{c}_{j}\hat{c}_{j+1}+\hat{c}_{j+1}^{\dagger}\hat{c}_{j}^{\dagger}\right].\end{split}\end{equation}
Here $\hat{c} _j\pap{\hat{c} _j^{\dagger}}$  is the annihilation (creation) operator of spinless fermions at site $j=1,\ldots, L$, $\mu$ is the chemical potential, $t$ is the hopping amplitude between nearest-neighbor sites (we assume $t\geq 0$ without loss of generality) and $\Delta$ is the nearest-neighbor superconducting induced pairing interaction. The Kitaev model features two phases: a topological and a trivial phase. In the former the Majorana ZEM emerge, which occurs whenever $|\mu| <\pm 2\Delta$ for the symmetric hopping-pairing Kitaev Hamiltonian, i.e., $t=\Delta$, the case we restrict ourselves from now on~\cite{majorana,Kitaev}. Additionally, the general interaction Hamiltonian is given by~\cite{MirceaTrifHamiltonian}
\begin{equation}
\hat{\mathcal{H}}_{{\rm Int}}=\pap{\frac{\hat{a}^\dagger+\hat{a}}{\sqrt{L}}}\bigg[\lambda_0\sum_{j=1}^L \hat{c}_j^\dagger \hat{c}_j+\frac{\lambda_1}{2}\sum_{j=1}^{L-1}\pap{\hat{c}_{j}^\dagger \hat{c}_{j+1}+ \hat{c}_{j+1}^{\dagger} \hat{c}_{j}}\bigg].
\end{equation}
Thus, for the light-matter coupling, we shall consider a general case which incorporates both on-site ($\lambda_0$) as well as hoppinglike ($\lambda_1$) terms (without loss of generality we will assume $\lambda_0$, $\lambda_1>0$). In Ref.~\cite{MirceaTrifHamiltonian}, a typical value of the on-site chain-cavity coupling, $\lambda_0 \simeq 0.1\omega$ was estimated for a fermion chain length of $L=100$ sites. Note that the whole chain is assumed to be coupled to the same cavity field.

\section{Mean-Field Approach}\label{Sec:MF}
In order to gain physical insights on how the original topological phase of the Kitaev chain is modified by its coupling to a cavity, we start by performing a Mean-Field (MF) treatment. Although we develop the MF analysis for a chain with periodic boundary conditions, the relations we will discuss in this section are indeed useful guides for interpreting the quasiexact results obtained by density matrix renormalization group (DMRG) numerical simulations in chains with open boundary conditions, as illustrated below.\\
\\
We start by separating the cavity and the chain subsystems by describing their interaction as the mean effect of one subsystem over the other. Applying the traditional MF approximation to the interaction Hamiltonian $\hat{\mathcal{H}}_{{\rm Int}}$, we set  quantum fluctuations of products of bosonic and fermionic operators to 0, therefore
\begin{equation}
\pap{\hat{a}^\dagger+\hat{a}-\langle\hat{a}^\dagger+\hat{a}\rangle}\pap{\hat{c}_j^\dagger \hat{c}_j-\langle\hat{c}_j^\dagger \hat{c}_j\rangle}=0.
\end{equation}
Following a similar procedure for the hoppinglike light-matter interaction term and setting periodic boundary conditions, the new interaction Hamiltonian is given by

\begin{equation}\label{InteractionBipartition}
\begin{split}
\hat{\mathcal{H}}_{{\rm Int}}^{{\rm MF}}\approx & L\pas{\lambda_1 D+\lambda_0(1- S_z)}\pas{\hat{X}-x}\\
&+2x\bigg[\lambda_0\sum_{j=1}^L \hat{c}_j^\dagger \hat{c}_j+\frac{\lambda_1}{2}\sum_{j=1}^{L}\pap{\hat{c}_{j}^\dagger \hat{c}_{j+1}+ \hat{c}_{j+1}^{\dagger} \hat{c}_{j}}\bigg].
\end{split}
\end{equation}
Here, we define 

\begin{equation}
\begin{split}
        \hat{X}&=\pap{\hat{a}+\hat{a}^{\dagger}}/2\sqrt{L}, \quad x=\langle \hat{X}\rangle, \\
          S_z&=1-\frac{2}{L}\sum_j\langle \hat{c}^\dagger_{j} \hat{c}_{j}\rangle,\\                D&=\frac{1}{L}\sum_j \langle \hat{c}^\dagger_{j} \hat{c}_{j+1}+\hat{c}^\dagger_{j+1}\hat{c}_{j}\rangle, \end{split} \end{equation}

where expectation values are taken with respect to the photon-fermion ground-state. The resulting Hamiltonian is that of a displaced harmonic oscillator, with photon number $\langle \hat{a}^\dagger \hat{a} \rangle\equiv \langle \hat{n} \rangle=L x^2$, and a Kitaev chain with effective chemical potential $\mu_{{\rm eff}}\equiv\mu -2\lambda_0 x$ and hopping interaction $ t_{{\rm eff}}\equiv \Delta-\lambda_1x$  (see Appendix \ref{App:MFA} for more details on this MF approach).\\

The minimization of the MF Hamiltonian expected value, $\partial \langle \hat{\mathcal{H}}_{{\rm MF}}\rangle/ \partial x=0$, yields
\begin{equation}
\label{Minimization}
 \lambda_0S_z=\lambda_0+\lambda_1D+2 \omega x,
\end{equation}
 which shows the interdependence of the cavity and chain states parameters. Since $x\in [-\frac{2\lambda_0+\lambda_1}{2 \omega}, 0]$, the effective MF renormalized Kitaev parameters turns out to be $\mu_{{\rm eff}} \geq \mu$ and $t_{{\rm eff}} \geq \Delta$. By choosing $\lambda_1=0$, it is easy to see that $x$ will be related to the magnetization in the equivalent transverse Ising chain~\cite{qpt,susuki,PolaritonLiberato,Ising-Kitaev}, while when choosing $\lambda_0=0$, $x$ will be associated to the occupancy of first neighbor nonlocal Majorana fermions in the Kitaev chain~\cite{FernandoTwoTimeCorrelations}.
\section{Phase Diagram}
\label{Sec:PhaseDiagram}
The ground-state of the system has been obtained by performing DMRG simulations in a matrix product state description~\cite{SCHOLLWOCK,Or_s_2019DMRG}, using the open-source TNT library~\cite{TNT,Sarah}. Notably, matrix product algorithms have been successfully applied to correlated systems embedded in a cavity~\cite{GandMDMRG,CavityKollath}, as well as to different interacting systems in starlike geometries~\cite{wolf2014prb,Mendoza:2017,zwolak2020prl,brenes2020}. In the following analysis, we consider separately each kind of cavity-chain coupling term and we sweep over $\mu$.\\
\\
The topological phase of the chain will be assessed through the two-end correlations $Q$, defined as $Q\equiv 2\langle \hat{c}_1 \hat{c}_L^\dagger+\hat{c}_L \hat{c}_1^\dagger \rangle$. This value is an indicator of the locality of edge modes, which is connected to the topology of the system \cite{ReslenQ,LeeQ} . For an infinite isolated Kitaev chain, its value is $1$ in the topological phase while it goes to $0$ in the trivial one. However, for finite sizes the value of $Q$ takes on continuous values in between, leaving a value of $1$ at the point of maximum correlations [cf.  Figs.~\ref{fig:Q_PD}(e) and \ref{fig:Q_PD}(f), see Appendix \ref{App:PhaseCharacterization}]. Whenever $Q>Q_{{\rm Trigger}}$ the phase is said to be topological, where $Q_{{\rm Trigger}}$ was defined as the lowest $Q$ that allows for ZEM to emerge in an isolated Kitaev chain with the same $\Delta$ and $L$ as in the simulated light-coupled case. In Appendix \ref{App:PhaseCharacterization}, we show the agreement of this definition of the topological phase with the description provided by a topological invariant, namely, the Majorana number \cite{Kitaev}.\\

 For both types of cavity couplings, second-order phase transitions arise in the composite light-matter model (an example is shown in Appendix \ref{App.DMRGVsMF}), a result for which DMRG and MF are in full agreement for a wide range of coupling values.
\begin{figure}[t!]
\begin{center}
\includegraphics[width=1\linewidth]{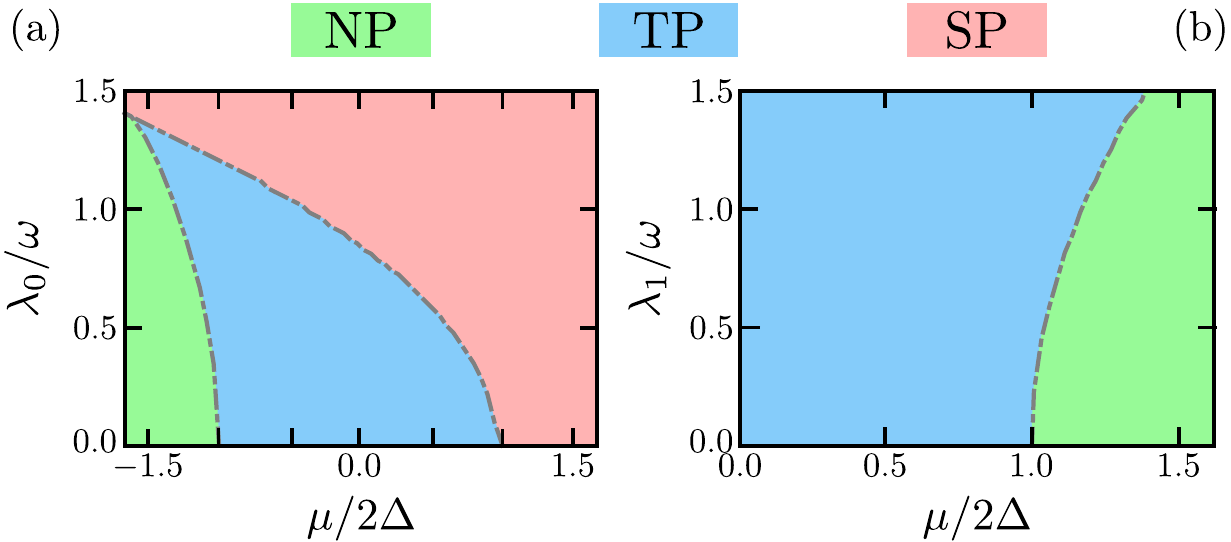}
\caption{Photon-fermion phase diagrams. NP: normal phase, TP: topological phase and SP: asymptotically super-radiant phase. \text{(a)} Chemical potential-like coupling. \text{(b)} hoppinglike coupling. The Kitaev-cavity parameters are $L=100$ and $\Delta =0.6 \omega$.}\label{fig:PhaseDiagram}
\end{center}
\end{figure}
The phase diagram for the on-site coupling ($\lambda_0 \neq 0$ and $\lambda_1 = 0$) is presented in Fig. \ref{fig:PhaseDiagram}(a), whereas that for the hoppinglike coupling ($\lambda_0 = 0$ and $\lambda_1 \neq 0$) is depicted in Fig.~\ref{fig:PhaseDiagram}(b).\\

For the on-site coupling, the critical points and the maximum of correlations move asymmetrically to lower values of the chemical potential as the coupling strength increases (see Appendix \ref{App:PhaseCharacterization}). The boundary between the topological phase (TP) and the asymptotically super-radiant phase (SP), in which the number of photons approaches the maximum obtained by MF [cf. Fig.~\ref{fig:NumberOfPhotons}(a), see Appendix \ref{App:PhaseCharacterization}], is affected more dramatically causing the TP to disappear beyond $\lambda_0/\omega =1.39\pm 0.01$. For larger values of $\lambda_0$, there will only be one second-order phase transition between  the normal phase (NP), which is topologically trivial and does not present radiation, and SP, holding only a trivial ordering of the chain.\\
\\
For the hoppinglike photon-chain coupling case, the phase transition points are symmetrical with respect to the transformation $\mu \rightarrow -\mu$ (see Appendix \ref{App.DMRGVsMF}). Whenever the cavity resides in a super-radiant phase, the chain is in the topological phase (see Appendix \ref{App:PhaseCharacterization}); thus the mean number of photons acts as an orderlike parameter that correlates well with the quantum state of the chain. It is evident that this type of cavity-chain coupling widens the topological phase allowed region. However, as the TP gets wider the maximum value of $Q$ decreases, indicating the degrading of nonlocal chain correlations at high coupling values.
\section{von Neumann entropy, criticality and Gaussian-states}
\label{Sec:VNCRiticalityGS}
\begin{figure}[t!]
\begin{center}
    \includegraphics[width=1\linewidth]{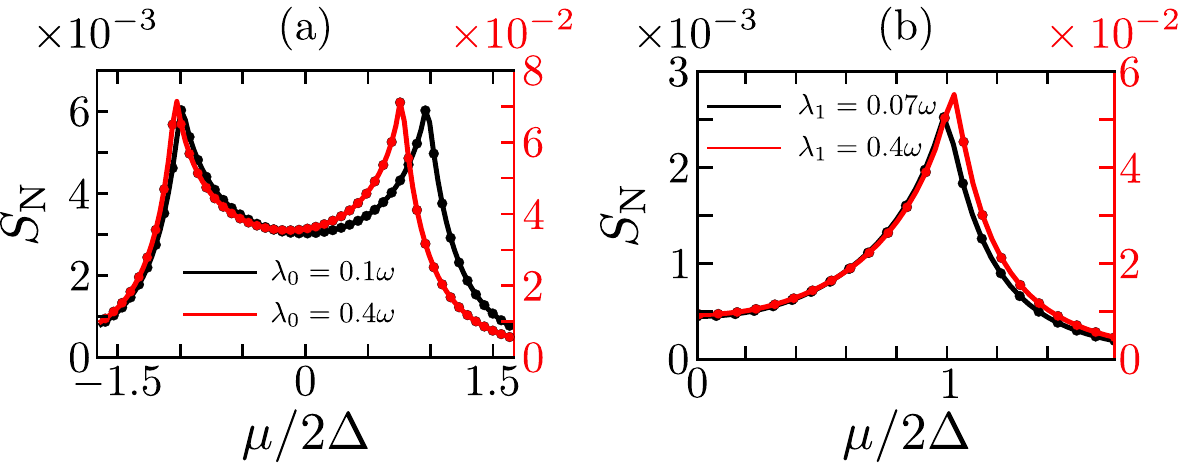}
\end{center}
 \caption{von Neumann entropy $S_{N}$ as a function of the chemical potential of the chain $\mu/2\Delta$ for any subsystem in the bipartite cavity-chain system. Symbols (lines) indicate DMRG (Gaussian) results. (a) Local photon-fermion couplings $\lambda_0 = 0.1\omega$ (weak-coupling, black symbols and line) and $\lambda_0 = 0.4\omega$ (moderate coupling, red symbols and line). (b) nonlocal photon-fermion coupling $\lambda_1 = 0.07\omega$ (weak-coupling, black symbols and line) and $\lambda_1 = 0.4\omega$ (moderate coupling, red symbols and line). Other parameters are $L=100$, $\omega= 1$ and $\Delta=0.6\omega$}
    \label{fig:Entropy}
\end{figure}

A result well beyond the MF analysis for this photon-fermion system is that phase transitions are associated with singularities in the light-matter quantum von Neumann entropy, $S_{{\rm N}}$~\cite{Vidal,Sarah,Lewis_SwanDM}, as shown in Fig.~\ref{fig:Entropy}. Critical lines, as obtained from the nonlocal $Q$-correlation behavior, are fully consistent with results extracted from the second derivative of the energy and $S_{{\rm N}}$ behavior (see. Appendix \ref{App.DMRGVsMF}). Moreover, the maximum nonlocal edge correlation $Q=1$ coincides with the minimum of $S_{{\rm N}}$ [compare Figs.~\ref{fig:Q_PD}(e) and \ref{fig:Q_PD}(f) with Fig.~\ref{fig:Entropy}(a)]. Consistency with $S_{{\rm N}}$ singularities is also found for the hoppinglike coupling, as shown in Fig.~\ref{fig:Entropy}(b). In any case, singularities in $S_{{\rm N}}$ are intimately connected to the phase transitions for an ample domain of coupling strength parameters (see also weak-coupling behaviors in Fig.~\ref{fig:Entropy} for $\lambda_0=0.1\omega$ and $\lambda_1=0.07\omega$).
\\

The MF analytical description, which involves a single coherent state for the cavity, provides an accurate description of the bulk expectation values in the chain, the mean number of cavity photons, the cavity quadratures and the energy of the whole system (see Appendix \ref{App.DMRGVsMF}). However, this effective theory is unable to account for entanglement properties between subsystems and higher interaction terms such as the FF. Many of the properties of GS have been broadly studied~\cite{GaussianExpectedOlivares, GaussianRenyiDaeKil, GaussianVonNeumann, GaussianAlexanian}, being one of the most outstanding the fact that it is fully characterized by its $2 \times 2$ covariance matrix $\sigma$ and first moments of the field-quadrature canonical variables given by $\hat{q} = \pap{\hat{a}^\dagger+\hat{a}}/\sqrt{2}$ and $\hat{p} = i \pap{\hat{a}^\dagger-\hat{a}}/\sqrt{2}$. The covariance matrix for a single-mode GS is simply
\begin{equation}
    \sigma =\begin{pmatrix}
    \langle \hat{q}^2 \rangle -\langle \hat{q} \rangle^2 &   \langle \hat{q}\hat{p} + \hat{p}\hat{q} \rangle -\langle \hat{q} \rangle \langle \hat{p} \rangle\\
        \langle \hat{q}\hat{p} + \hat{p}\hat{q} \rangle -\langle \hat{q} \rangle \langle \hat{p} \rangle & \langle  \hat{p}^2 \rangle -\langle \hat{p} \rangle
   \end{pmatrix}.
\end{equation}
\\

Remarkably, an accurate description of the reduced photon system density matrix is possible by means of a single mode GS. Any single-mode GS can be expressed in terms of a fictional thermal state on which squeezed ($\hat{S}_\xi$) and displacement ($\hat{D}_\alpha$) operators act in the form:
\begin{equation}
  \hat{\rho}_{\rm GS}=\hat{D}_\alpha \hat{S}_\xi\frac{N ^{\hat{a}^\dagger \hat{a}}}{\pap{1+N}^{a^\dagger a}}S^\dagger_\xi D^\dagger_\alpha,
\end{equation}
with the operators defined as 
\begin{equation}
\begin{split}
        \hat{D}_\alpha&=\exp\pas{\alpha \hat{a}^\dagger - \alpha^* \hat{a}}, \\  \hat{S}_\xi&=\exp\pas{\pap{ \xi^* (\hat{a})^2 - \xi (\hat{a}^\dagger)^2 }/2},
\end{split}
\end{equation}
 where $\alpha\in \mathbb{C}$, $\xi=re^{i\phi}$ is an arbitrary complex number with modulus $r$ and argument $\phi$,  and $N$ is the thermal state parameter~\cite{GaussianExpectedOlivares}. Furthermore, a well known property is that $S_{{\rm N}}$ is maximized for a single-mode GS at given quadrature variances and it is simply expressed as~\cite{GaussianVonNeumann, GaussianAlexanianExpected, GaussianRenyiDaeKil}
 \begin{equation}
 \label{Eq:GaussianSN}
     S_{{\rm N}}=\pap{N+1}\ln\pas{N+1}-N\ln\pas{N}.
 \end{equation}
 
In order to get the $\alpha$, $N$, $r$, and $\phi$ Gaussian parameters, the covariance matrix and quadratures are numerically extracted from the corresponding expected values using ground-state DMRG calculations. The imaginary part of $\alpha$ and $\phi$ must be $0$ to reach the ground-state. Therefore, the relations that endorse us with the Gaussian parameters are the following~\cite{GaussianExpectedOlivares}:
\begin{subequations}

\begin{align}
  \langle \hat{q} \rangle&=\sqrt{2} \Re\pas{\alpha},\\
   \langle \hat{p} \rangle&=\sqrt{2} \Im\pas{\alpha}\\
   \langle \hat{q}^2 \rangle -\langle \hat{q} \rangle^2=\frac{1+2N}{2}&\pap{\cosh\pas{2r}+\sinh\pas{2r}\cos\pas{\phi} },\\
  \langle  \hat{p}^2 \rangle -\langle \hat{p} \rangle^2=\frac{1+2N}{2}&\pap{\cosh\pas{2r}-\sinh\pas{2r}\cos\pas{\phi}},\\
  \langle \hat{q}\hat{p} + \hat{p}\hat{q} \rangle -\langle \hat{q} \rangle \langle \hat{p} \rangle&=\frac{1+2N}{2}\pap{\sinh\pas{2r}\sin\pas{\phi} }.
\end{align}
\end{subequations}
 Results for $S_{{\rm N}}$ obtained from DMRG and analytical GS calculations of Eq. \eqref{Eq:GaussianSN} are in excellent agreement for different coupling types and strengths, as shown in Fig. \ref{fig:Entropy}, thus confirming the adequacy of a GS photon description for the present photon-fermion system (see also Appendix \ref{App:GaussianParameters}).
%==================== SECTION =====================
\section{R\'enyi entropy and Fano factor}
\label{Sec:Renyi}
The R\'enyi entropies, defined as
\begin{equation}
    S_{\eta}\pap{\hat{\rho}}=\pap{1-\eta}^{-1}\ln\pas{\tr \pas{\hat{\rho}^{\eta}}}
\end{equation}
for a state $\hat{\rho}$, have been identified as powerful indicators of quantum correlations in multipartite systems~\cite{adessoPRL2012}. The von Neumann entropy $S_{{\rm N}}$ is retrieved as the R\'enyi entropy in the limit $\eta \rightarrow 1$. It has also been established that the R\'enyi entropy of order $\eta=2$ is well adapted for extracting correlation information from GS. Thus, from now on we restrict ourselves to consider only $S_{2}(\rho)=-\ln\pas{\tr (\rho^{2})}$ which we will simply denote as $S_{{\rm R}}$ ~\cite{Lewis_SwanDM, REHubbard&Spin,EntanglementInManyBody}. Specifically, $S_{{\rm R}}$ for a GS can be simply expressed in terms of the GS covariance matrix $\sigma$ as $S_{{\rm R}}=\frac{1}{2}\ln\pas{\det (\sigma)}$ \cite{GaussianRenyiDaeKil}.\\
\\
We also consider the photon FF, which is defined as FF$ = {\rm Var}\pap{\hat{n}}/\langle \hat{n}\rangle$, with ${\rm Var}\pap{\hat{n}}=\langle \hat{n}^{2} \rangle - \langle \hat{n}\rangle^{2} $. For further reference, FF$=1$ for a single coherent state (MF result) while it denotes either a sub- (FF$<1$) or super- (FF$>1$) Poissonian photon state. We now argue that the GS approximation allows us to analytically work out a relation between the FF and the entanglement entropy $S_{\rm R}$, raising them as both reliable and accessible indicators of phase transitions in composed photon-fermion systems. For a cavity GS, the FF and the $S_{\rm R}$ can be analytically expressed as~\cite{GaussianExpectedOlivares,GaussianAlexanianExpected,GaussianRenyiDaeKil}:
\begin{figure}
\begin{center}
 \includegraphics[width=1.0\linewidth]{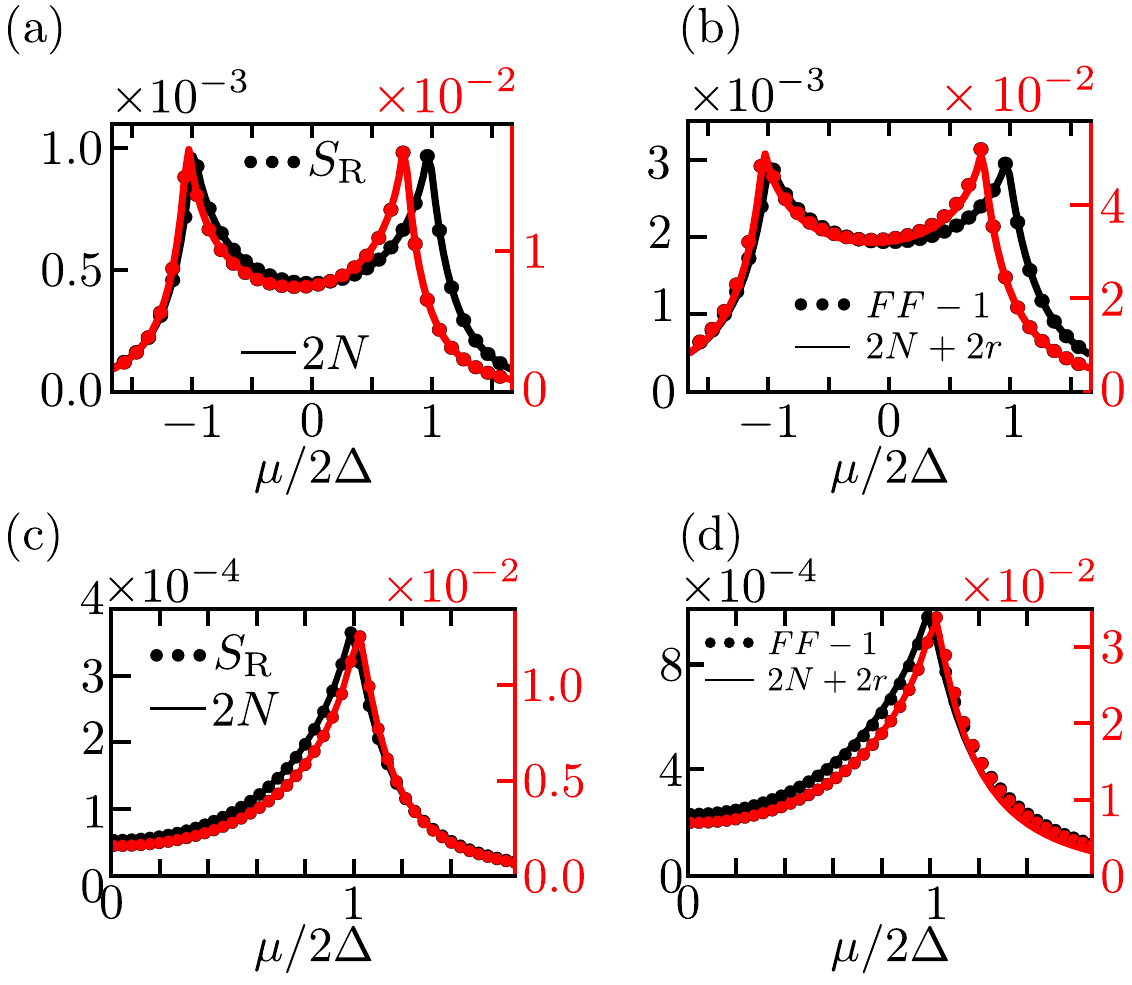}
\end{center}
  \caption{[(a) and (c)] R\'enyi entropy $S_{\rm R}$ and [(b) and (d)] Fano factor (FF$-1$) of the cavity state as a function of the chemical potential of the chain $\mu/2\Delta$. Symbols (lines) indicate DMRG (GS) results. [(a) and (b)] Results for local photon-fermion coupling, $\lambda_0=0.1\omega$ (black) and $\lambda_0=0.4\omega$ (red). [(c) and (d)] Results for nonlocal photon-fermion coupling, $\lambda_1=0.07\omega$ (black) and $\lambda_1=0.4\omega$ (red). Other parameters are $L=100$, $\omega=1$, and $\Delta=0.6\omega$.}
  \label{fig:GaussianVariables}
\end{figure}
\begin{align}
    {\rm FF}&=\frac{(N+1/2)^2\cosh\pas{4r}+(1+2N)e^{2r}\alpha^2-1/2}{(N+1/2)\cosh\pas{2r}+\alpha^2-1/2},\label{eq:GaussianFano}\\
    S_{\rm R}&=2\ln\pas{1+N}+\ln\left[1-\left(\frac{N}{1+N}\right)^2\right].\label{eq:GaussianRenyi}
\end{align}

For both kinds of photon-fermion couplings, results obtained from these analytical expressions fit exactly the numerical ones extracted from full DMRG calculations. Assuming a GS, the inequalities $N, r \ll \abs{\alpha}$ and  $N,\, r \ll 1$, which allow to clearly see the connection between both quantities, are reliable and well justified for the range of parameters of experimental interest (see Appendix \ref{App:GaussianParameters}). Keeping first order terms in $r$ and $N$, in Eqs.~\eqref{eq:GaussianFano} and~\eqref{eq:GaussianRenyi}, we finally get FF$=1+2(r+N)$ (i.e., a super-Poissonian photon state) and $S_{\rm R}=2N$, from which a simple relation between $S_{\rm R}$, FF and the squeezing parameter $r$ immediately follows as
\begin{equation}
\label{eq:Fano&Renyi}
S_{\rm R}={\rm FF}-2r-1.
\end{equation}
It must be stressed that this last relation between the Rényi entropy and cavity observables does not rely on a MF analysis or equivalently on the assumption of a single coherent state for the cavity, but rather it comes exclusively from numerical DMRG calculations and their analytical backing by a Gaussian-state approximation.\\

\begin{figure}
    \centering
    \includegraphics[width=1.0\linewidth]{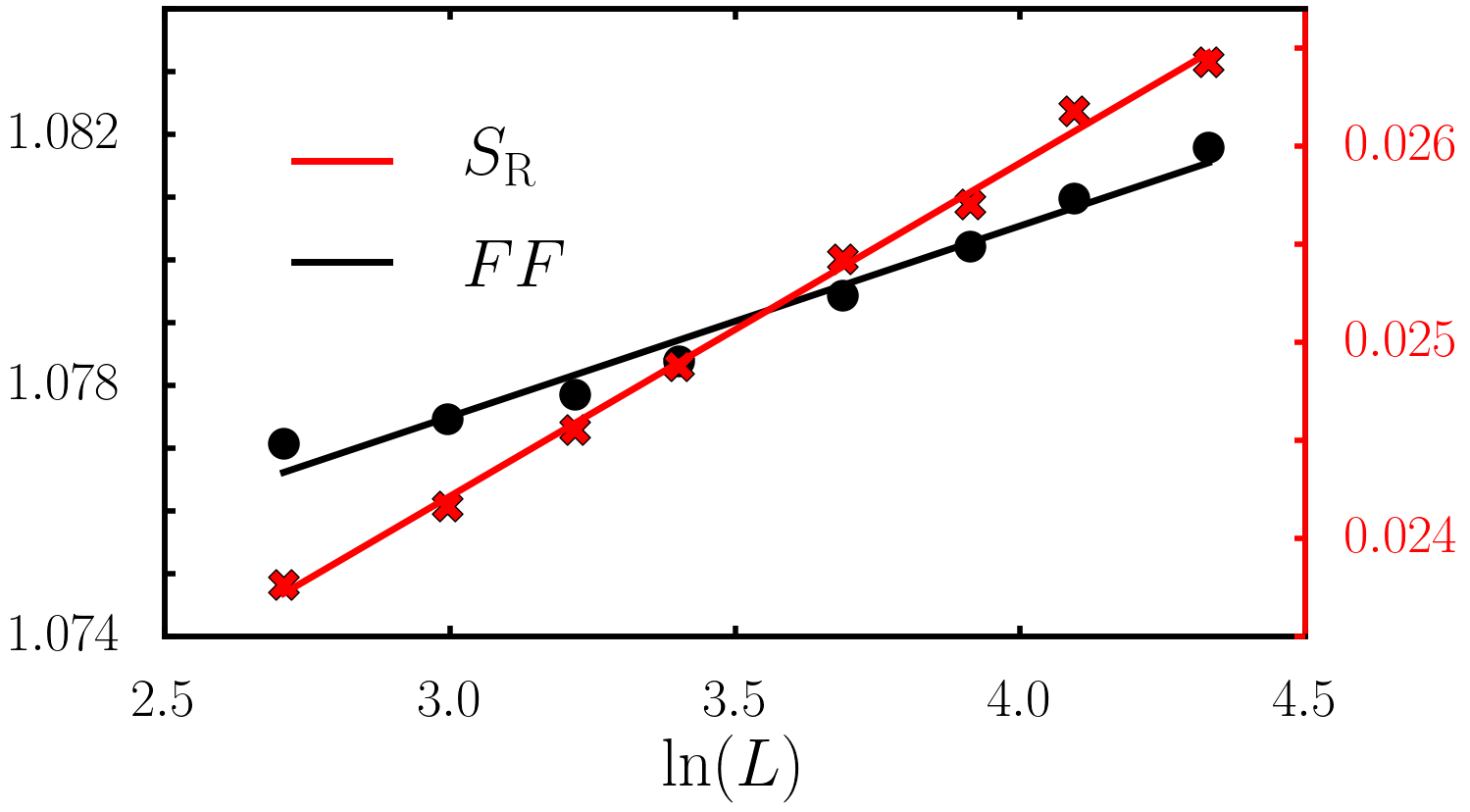}
    \caption{Scaling of FF (black symbols and left scale) and $S_{\rm R}$ (red symbols and right scale) with the lattice size at the critical point between the TP and SP for the on-site coupling. The correspondent colored solid lines are the result of a linear fit, which is the typical form of the growth of entanglement at criticality for one-dimensional systems. The parameters are $\Delta=0.6\omega$ and $\lambda_0=0.49 \omega$. }
    \label{fig:FF&RECriticality}
\end{figure}

The validity of this important result is illustrated in Fig.~\ref{fig:GaussianVariables} regardless of the photon-fermion coupling type. In spite of the similar behavior through a topological phase transition (and corresponding analytical expressions for a GS) of von Neumann and R\'enyi entropies, it is important to note that an equivalent relation to that in Eq.~\eqref{eq:Fano&Renyi} but involving $S_{{\rm N}}$ instead of $S_{\rm R}$ is hardly workable. Therefore, we stress the relevance of this connection between a theoretical quantum information entropy, $S_{\rm R}$, and measurable photon field observables, FF and $r$.\\
\\
Figures~\ref{fig:GaussianVariables}(a) and \ref{fig:GaussianVariables}(b) exhibit the behavior of different terms involved in Eq.~\eqref{eq:Fano&Renyi} for the local photon-fermion coupling ($\lambda_0=0.1 \omega$ and $0.4 \omega$), and show an excellent agreement between the results directly obtained from DMRG and those assuming a cavity GS. This validates Eq.~\eqref{eq:Fano&Renyi}, according to which $S_{\rm R}+2r$ and FF$-1$ coincide. Very small deviations between GS and DMRG results at the topological phase transition are observed, for the stronger coupling value. However, the locations of the singularities predicted by the analytical and numerical results coincide. Similarly, Figs.~\ref{fig:GaussianVariables}(c) and \ref{fig:GaussianVariables}(d) display respective calculations for a hoppinglike coupled system ($\lambda_1=0.07$ and $0.4$), showing that GS results seem to slightly drift apart from the numerically exact DMRG ones for the highest coupling.\\
\\
We observe that the squeezing gets larger as the light-matter subsystems become more entangled at the critical point (note the behavior of the $r$ parameter comparing the different curves in Fig.~\ref{fig:GaussianVariables},  see also Appendix \ref{App:GaussianParameters}). In order to measure the squeezing parameter $r$, one can resort to a homodyne detection technique which has been recently extended to the microwave spectral region~\cite{Haus_2004squeezed,andrews2015photonics,2modeSqueezingWallraff}. On the other hand, the FF can be assessed from measurements of the second-order correlation function $g^{(2)}$(for the relation between both see Refs. \cite{GaussianExpectedOlivares,GaussianAlexanianExpected}), which has successfully been measured in experiments \cite{LangG2}. Moreover, in Fig. \ref{fig:FF&RECriticality} we show a scaling analysis on the value of both, FF and $S_{\rm R}$, at criticality. The results reveals a logarithmic growth with the size of the chain, which is reminiscent of the behavior of entanglement at criticality in 1D systems \cite{EntanglementAtCriticality}. From the results of Fig. \ref{fig:FF&RECriticality} together with Eq. \eqref{eq:Fano&Renyi} it is straightforward to see that the squeezing at criticality depends logarithmically with the the size of the chain. This behavior is not unique to the hoppinglike coupling, but we found this logarithmic response of the cavity for the chemical potential-like interaction as well.  Thus, the FF behavior and its very close relation to $S_{\rm R}$ turn out to be measurable, reliable and accurate indicators of entanglement for this light-matter interacting system.  Aside from the fact that it is always interesting to establish the connections between different approaches, our main result in Eq.~\eqref{eq:Fano&Renyi} raises the question of whether a GS approximation remains valid for quantum open systems and/or stronger light-matter coupling strengths. For example, photon loss from the cavity is a ubiquitous deleterious effect in experimental setups, but key to measure the state of the cavity field. These subjects merit considerably further studies, motivated by our work.

\section{Conclusions}
\label{Sec:Conclusions}
In this work, we have developed a direct link between accessible microwave observables and quantum entanglement entropies in quantum matter featuring topological phase transitions. By resorting to a GS description for the photon subsystem, as supported by DMRG calculations, we found a simple but powerful relation between the photon Fano factor, single-mode quadrature amplitudes and the light-matter R\'enyi entropy. Singularities in the latter can then be of help for characterizing topological phase transitions and their connection to nonmonotonic nonlocal correlations in a fermionic chain. We also provide evidence of how the topological phase can be modified with both on-site as well as hopping terms of photon-fermion interactions, yielding in some cases to a more robust topological phase. The possibility of extracting nonlocal or topological information of the Kitaev chain from the photonic field itself should be highly timely given the continuous challenges to assess in a clean way Majorana features in transport experiments. Moreover, our results also open novel questions which motivate further studies of the role of decoherence on this quantum light-matter system.\\
\begin{acknowledgments}
The authors acknowledge the use of the Universidad de los Andes High Performance Computing (HPC) facility in carrying out this work. J.J.M-A, F.J.R and L.Q are thankful for the support of MINCIENCIAS, through the Project ``{\it Producci\'on y Caracterizaci\'on de Nuevos Materiales Cu\'anticos de Baja Dimensionalidad: Criticalidad Cu\'antica y Transiciones de Fase Electr\'onicas}" (Grant No. 120480863414) and from Facultad de Ciencias-UniAndes, projects ``{\it Quantum thermalization and optimal control in many-body systems}" (2019-2020) and ``{\it Excited State Quantum Phase Transitions in Driven Models - Part II: Dynamical QPT}" (2019-2020). C.T acknowledges financial support from the Ministerio de Econom\'ia y Competitividad (MINECO), under Project No. MAT2017-83772-R. F.J.G-R acknowledge funding support from the {\it Ministerio de Ciencia e Innovaci\'on}, under Project No. PID2019-109007GA-I00.
\end{acknowledgments}
%====================== APPENDIX =============================
\appendix
\section{Matrix product operator representation}
To find the ground-state of the fermion-photon system with DMRG, it is necessary to find a way to write the Hamiltonian in Eq.~\eqref{FullHamiltonian} in a matrix product operator (MPO) representation. The latter can be interpreted as describing any system operator of interest as a product of matrices that only contains operators of a single site, in a 1D arrangement of the system. For instance, a Hamiltonian $\hat{\mathcal{H}}$ acting over a 1D lattice with $L$ sites is required to be represented as $\hat{\mathcal{H}}=\prod^{L}_{i=1} W^i$, where $W^i$ is a matrix that only contains operators of site $i$. \\
\\
The Hamiltonian we consider here describes a chain, with nearest-neighbor interactions, coupled to a global site [a single cavity field in the case of the Hamiltonian Eq. \eqref{FullHamiltonian}], thus having a starlike geometry. The cavity field is assumed to interact with the whole chain. The Hamiltonian can thus be written in the following way:

\begin{equation}\label{InteractionBipartition}
\begin{split}
        \hat{\mathcal{H}}=\sum^{L}_{i=1}&\hat{h}_i+\sum^{\alpha}_{k=1}\sum^{L-1}_{i=1}\hat{m}_i^k\hat{n}_{i+1}^k+\sum^{\beta}_{k=1}\hat{A}^k\sum^{L-1}_{i=1}\hat{x}_i^k\hat{y}_{i+1}^k\\
        +&\sum^{\gamma}_{k=1}\hat{B}^k\sum^{L}_{i=1}\hat{z}_i^k+\hat{C},
\end{split}
\end{equation}

where $L$ is the size of the chain. Here, $\hat{A}^k$, $\hat{B}^k$, and $\hat{C}$ are operators that act on the global site (cavity). On the other hand, $\hat{h}_i$, $\hat{m}_i^k$, $\hat{n}_i^k$, $\hat{x}_i^k$, $\hat{y}_i^k$ and $\hat{z}_i^k$ are single-site chain operators. Additionally, $\alpha$, $\beta$, and $\gamma$ denote, respectively, the minimum number of operators needed to conform terms corresponding to nearest-neighbor interactions in the chain ($\Delta$); nearest-neighbor interactions in the chain with the global site ($\lambda_1$); and on-site chain terms coupled with the cavity ($\lambda_0$). It is important to note that in this way we resort to a generalized 1D system which consists of $L+1$ sites, where the global site is located at the left edge of the 1D arrangement. Denoting the site $0$ as the global site, we designed an MPO for this type of Hamiltonian with matrices $W^i$ defined as follows:
  \begin{itemize}
   \item For site $L$, $W^L_a=W^L_{a,1}$, the matrix will be the first column vector of the corresponding matrix for the bulk.
    \item For $i \in [1,L-1]$,  $W^i \in T(d\times d)$ with $d=2+2\beta+\gamma+\alpha$ and $T(m\times n)$ the set of tensors with size $m\times n$, with elements
    \begin{equation*}
    W_{n,m}^{i}=\begin{cases}
    \hat{1}&\text{if } n=m=a, \text{ with }\\[0.6ex]
    &a\in\{1,d\} \cup \{b | b=2k+1, \: k\in[1,\beta] \}\\[0.6ex]
    & \cup \{c | c=k+2\beta+1, \: k\in[1,\gamma] \},\\
    &\\
    \hat{y}^k_i& \text{if } n=2k,\:m=1, \text{ with } k \in[1,\beta],\\
    &\\
    \hat{x}^k_i&\text{if } n=2k+1,\:m=2k, \text{ with } k \in[1,\beta],\\
    &\\
    \hat{z}^k_i&\text{if } n=2\beta+k+1,\:m=1, \text{ with } k \in[1,\gamma],\\
    &\\
    \hat{n}_{i}^k&\text{if } n=2\beta+\gamma+k+1,\:m=1\text{ with } k \in[1,\alpha],\\
    &\\
    \hat{m}_{i}^k&\text{if } n=d,\: m=2\beta+\gamma+k+1\text{ with } k \in[1,\alpha],\\
    &\\
    \hat{h}_i& \text{if }n=d,\:m=1,\\
    &\\
    0& \text{otherwise}.
    \end{cases}
    \end{equation*}
\item For the global site, i.e., site $i=0$, we have a row vector with the form:
   \begin{equation*}
   W_{m}^{0}=\begin{cases}
   \hat{C}& \text{if }  m=1\\
   &\\
   \hat{A}^k& \text{if }  m=2k+1,\; k\in\pas{1,\beta}\\
   &\\
  \hat{B}^k& \text{if } m=2\beta+k+1 ,\; k\in\pas{1,\beta}\\
  &\\
   \hat{1}& \text{if }  m=d\\
   &\\
   0 &\text{otherwise.}
   \end{cases}
   \end{equation*}
\end{itemize}In this way, the Hamiltonian of Eq.~\eqref{FullHamiltonian} is represented with the following MPO under a Jordan-Wigner transformation:

\begin{align}
   W^{i\in\pas{2,L}}&=
   \begin{pmatrix}
    \hat{1} & 0 & 0 & 0 & 0 & 0 & 0 &0\\
    \hat{\sigma}_x   & 0 & 0 & 0 & 0 & 0 & 0& 0\\
  0  & \frac{\lambda_1}{4\sqrt{L}}\hat{\sigma}_x  & \hat{1} & 0 & 0 & 0 & 0& 0\\
  \hat{\sigma}_y  & 0 & 0 & 0 & 0 & 0 & 0& 0\\
0 & 0 &\frac{\lambda_1}{4\sqrt{L}}\hat{\sigma}_y  & \hat{1} & 0 & 0 & 0& 0\\
    \frac{\lambda_0}{2\sqrt{L}}\hat{\sigma}_z  & 0 & 0 & 0 & 0 & \hat{1} & 0& 0\\
   \hat{\sigma}_y  & 0 & 0 & 0 & 0 & 0 & 0& 0\\
      \frac{\mu}{2}\hat{\sigma}_z   & 0 & 0 & 0 & 0& 0&-\Delta\hat{\sigma}_y  & \hat{1}
   \end{pmatrix},\\
   W^{L}&=\begin{pmatrix}
   \hat{1} \\
   \hat{\sigma}_x \\
   0\\
   \hat{\sigma}_y  \\
   0\\
   \frac{\lambda_0}{2\sqrt{L}}\hat{\sigma}_z  \\
   \sigma_y  \\
   \frac{\mu}{2}\sigma_z
   \end{pmatrix},\\[1.5ex]
   W^0=&\begin{pmatrix}
    \omega \hat{a}^\dagger  \hat{a}+\lambda_0 L\hat{X} & 0 &2\sqrt{L}\hat{X} &0 &2\sqrt{L}\hat{X} &2\sqrt{L}\hat{X} &0& \hat{1}
    \end{pmatrix},
\end{align}
where $\hat{\sigma}_x $,  $\hat{\sigma}_y $, and  $\hat{\sigma}_z $ are the Pauli matrices.
\section{Mean-field Approach}
\label{App:MFA}

In Sec. \ref{Sec:MF} we obtained the MF interaction term. This allows us to write the Hamiltonian, Eq.~\eqref{FullHamiltonian}, as the contribution of two independent systems corresponding to a Kitaev chain (in terms of fermionic operators) and a forced harmonic oscillator (bosonic operators), plus a constant energy. Then the MF Hamiltonian can be written as $\hat{\mathcal{H}}_{{\rm MF}}\approx \hat{\mathcal{H}}_{{\rm C}}^{{\rm MF}}+\hat{\mathcal{H}}_{{\rm K}}^{{\rm MF}}+\hat{\mathcal{H}}_{{\rm Const}}^{{\rm MF}}$. The MF Hamiltonians are defined as

\begin{subequations}
\begin{equation}
\hat{\mathcal{H}}_{\rm C}^{{\rm MF}} = \omega \hat{a}^\dagger \hat{a}+L\pas{\lambda_1 D+\lambda_0(1- S_z)}\hat{X},
\end{equation}
\begin{equation}
\begin{split}
\hat{\mathcal{H}}_{\rm K}^{{\rm MF}}= & -\pas{\frac{\mu}{2} -\lambda_0 x}\sum_{j=1}^L\left[  2\hat{c}_j^\dagger \hat{c}_j-\hat{1} \right]\\
& -\pas{t-\lambda_1x}\sum_{j=1}^{L}\left[\hat{c}_{j}^{\dagger}\hat{c}_{j+1}+\hat{c}_{j+1}^{\dagger}\hat{c}_{j}\right]\\
& +\Delta \sum_{j=1}^{L}\left[\hat{c}_{j}\hat{c}_{j+1}+\hat{c}_{j+1}^{\dagger}\hat{c}_{j}^{\dagger}\right],
\end{split}
\end{equation}
\begin{equation}
\hat{\mathcal{H}}^{{\rm MF}}_{\rm Const}=  L\pap{\lambda_0 S_z-\lambda_1 D}x.
\end{equation}
\end{subequations}
Thus, the eigenstates of the Hamiltonian are just products of chain and cavity states. With the MF Hamiltonian $\hat{\mathcal{H}}_{{\rm MF}}$ being identified, we proceed to describe the thermodynamics of the composed system (at finite temperature $T$, which later on will tend to zero) by simply replacing new effective parameters in the Kitaev Hamiltonian. The coupling with the cavity produces a displacement of both the chemical potential and the hopping terms in the form $\mu\rightarrow\mu_{{\rm eff}}\equiv\mu -2\lambda_0 x$ and $t\rightarrow t_{{\rm eff}}\equiv t-\lambda_1x$, thus defining $\hat{\mathcal{H}}_{{\rm K}}^{{\rm MF}}$. The Hamiltonian $\hat{\mathcal{H}}_{{\rm C}}^{{\rm MF}}$ and is the sum of all bosonic terms, and $\hat{\mathcal{H}}_{{\rm Const}}^{{\rm MF}}$ consists of the remaining constant terms.\\

\begin{figure}[t!]
 \includegraphics[width=0.9\linewidth]{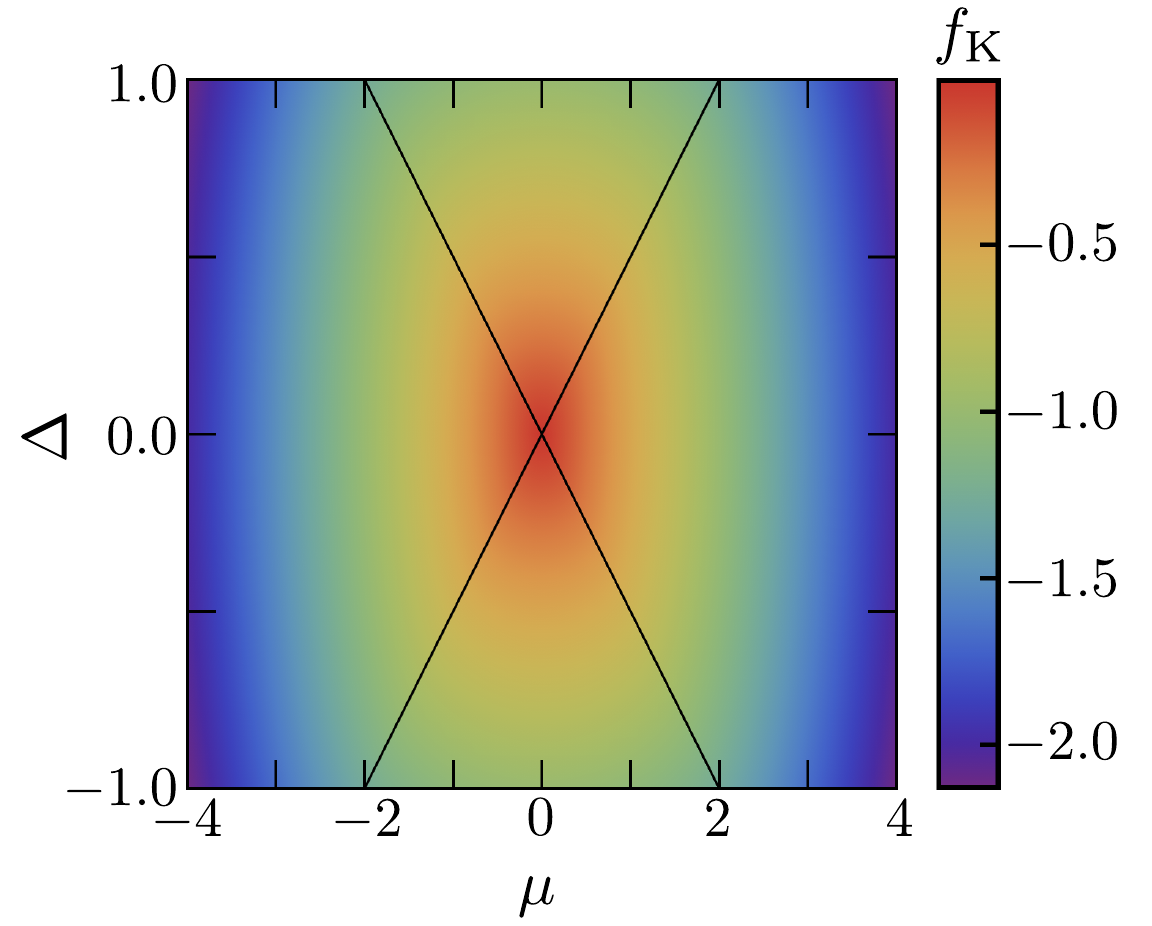}
		\caption{\label{fig:FreeEnergyKitaev} Density plot of the Kitaev mean-field free energy, $f_{\rm K}\equiv f_{\rm K}(x=0)$ with $\beta=100$ as a function of $\mu$ and $\Delta$. The black diagonal lines mark down the phase transition $\mu= \pm 2\Delta$.}
\end{figure}

The canonical partition function is given by
\begin{equation}
    \mathcal{Z}=\tr\pas{\exp\pas{-\beta\hat{\mathcal{H}}_{{\rm MF}}}},
\end{equation}
with $\beta=1/(k_{\rm B}T)$ and $k_{\rm B}$ the Boltzmann constant, is the product of three different terms, namely $\mathcal{Z}=\mathcal{Z}_{\rm C}*\mathcal{Z}_{\rm K}*\mathcal{Z}_{\rm Const}$
Following the common procedure to diagonalize the Kitaev Hamiltonian through a Bogoliubov-de Gennes quasiparticle description~\cite{susuki,majoranareturns,G&MMeanField}, we find
\begin{equation}
\hat{\mathcal{H}}_{{\rm K}}^{{\rm MF}}\pap{x}=\sum_k 2\omega_k\pap{x}\pap{ \hat{d}_k^\dagger \hat{d}_{k}-\frac{1}{2}},
\end{equation}
 where $\hat{d}_k$ ($\hat{d}_k^\dagger$) is the annihilation (creation) operator of Bogoliubov quasiparticles in momentum space $k$ at the first Brillouin zone. The dispersion relation is
 \begin{equation}
\omega_k(x)=\sqrt{\left[t_{{\rm eff}}\cos(k)+\mu_{{\rm eff}}/2\right]^2+\Delta^2\sin^2(k)}.
\end{equation}
 This results in the chain partition function $\mathcal{Z}_{\rm K}=\prod_k2\cosh(\beta \omega_k)$. The cavity term can be diagonalized by the displacement of the bosonic field, from which it is straightforward to obtain the partition function for the cavity term as well, given by 
 \begin{equation}
     \mathcal{Z}_{{\rm C}}=\frac{\exp\pas{\beta \phi^2/\omega}}{\pap{1-\exp\pas{-\beta \omega}}},
 \end{equation}
with $\phi=\sqrt{L}\pas{\lambda_1D+\lambda_0(1-S_z)}/2$. For the constant term, the effect in the partition function is trivial, namely
\begin{equation}
    \mathcal{Z}_{{\rm Const}}=\exp\pas{-\beta L\pap{\lambda_0 S_z-\lambda_1 D}x}.
\end{equation}
 Following the product form of $\mathcal{Z}$, the free energy, defined as $\mathcal{F}=-\ln\pas{\mathcal{Z}}/\beta$, is given by the addition of three terms: $\mathcal{F}=\mathcal{F}_{\rm K}+\mathcal{F}_{\rm C}+\mathcal{F}_{\rm Const}$. The free energy $\mathcal{F}$ thus reads
\begin{equation}
\label{eq:FreeEnergyNoReplacement}
    \mathcal{F}=\mathcal{F}_{\rm K}+\frac{1}{\beta L}\ln\pas{1-e^{-\beta \omega}}-\frac{\phi^2}{\omega}+ L\pap{\lambda_0 S_z-\lambda_1 D}x.
\end{equation}
Therefore, Eq. \eqref{Minimization} can be recovered from the free energy when performing the minimization $ \partial \mathcal{F}/ \partial S_z=0$.

By direct substitution of Eq.~\eqref{Minimization} on Eq.~\eqref{eq:FreeEnergyNoReplacement}, the free energy per site, $f\equiv \mathcal{F}/L$, results in an expression that only depends on the cavity expected value $x$:
\begin{equation}\label{eq:FreeEnergy}
        f(x)=f_{\rm K}(x)+\omega x^2 +\lambda_0x+\frac{1}{\beta L}\ln\pas{1-e^{-\beta \omega}},
\end{equation}
where the photonic part of the ground-state will be defined by the $x$ value that minimizes the free energy. The state of the cavity will be represented by a single coherent state $\ket{x\sqrt{L}}$. Note that this coherent state label does not have any imaginary part. The reason for this is that only the position quadrature explicitly appears in the Hamiltonian ($\hat{X}$ term, see Ref.~\cite{QuantumOpticsScully}). The momentum quadrature, which is directly related to the imaginary part of a coherent state (see Ref.~\cite{QFT}), is only implicitly regarded in the mean number of photons, $\langle \hat{a}^\dagger \hat{a} \rangle$, that appears in the Hamiltonian in Eq. \eqref{FullHamiltonian}. Since the effect of the imaginary part of a coherent state upon which the Hamiltonian of Eq. \eqref{FullHamiltonian} acts  is to add energy to the system, it sets the momentum quadrature to zero. Consequently, the mean number of photons satisfies $\langle \hat{a}^\dagger \hat{a} \rangle\equiv \langle \hat{n} \rangle=L x^2$. Last, the free energy of the Kitaev term reads
\begin{equation}\label{eq:KitaevFreeEnergy}
    f_{\rm K}(x)=-\frac{1}{\beta L}\sum_k  \ln\pas{2\cosh\pas{\beta \omega_k(x)}},
\end{equation}
which is shown in Fig. \ref{fig:FreeEnergyKitaev}. Then, when the system supports super-radiance in the ground-state, the free energy must meet the following condition for a given $x\neq 0$:
\begin{equation}
        f_{\rm K}(x)+\omega x^2+\lambda_0x-f_{\rm K}(0) < 0.
\end{equation}
A state with an expected value $x$ that satisfies the previous inequality will be less energetic than a state with no radiation in the composite system. This leads to a redefinition of the ground-state compared to the case of the isolated Kitaev chain. The latter means that the state of the cavity controls the free energy of the Kitaev chain by creating effective displacements, which are characterized by super-radiance in the cavity.

\begin{figure}[h!]
    \centering
    \includegraphics[width=1.0\linewidth]{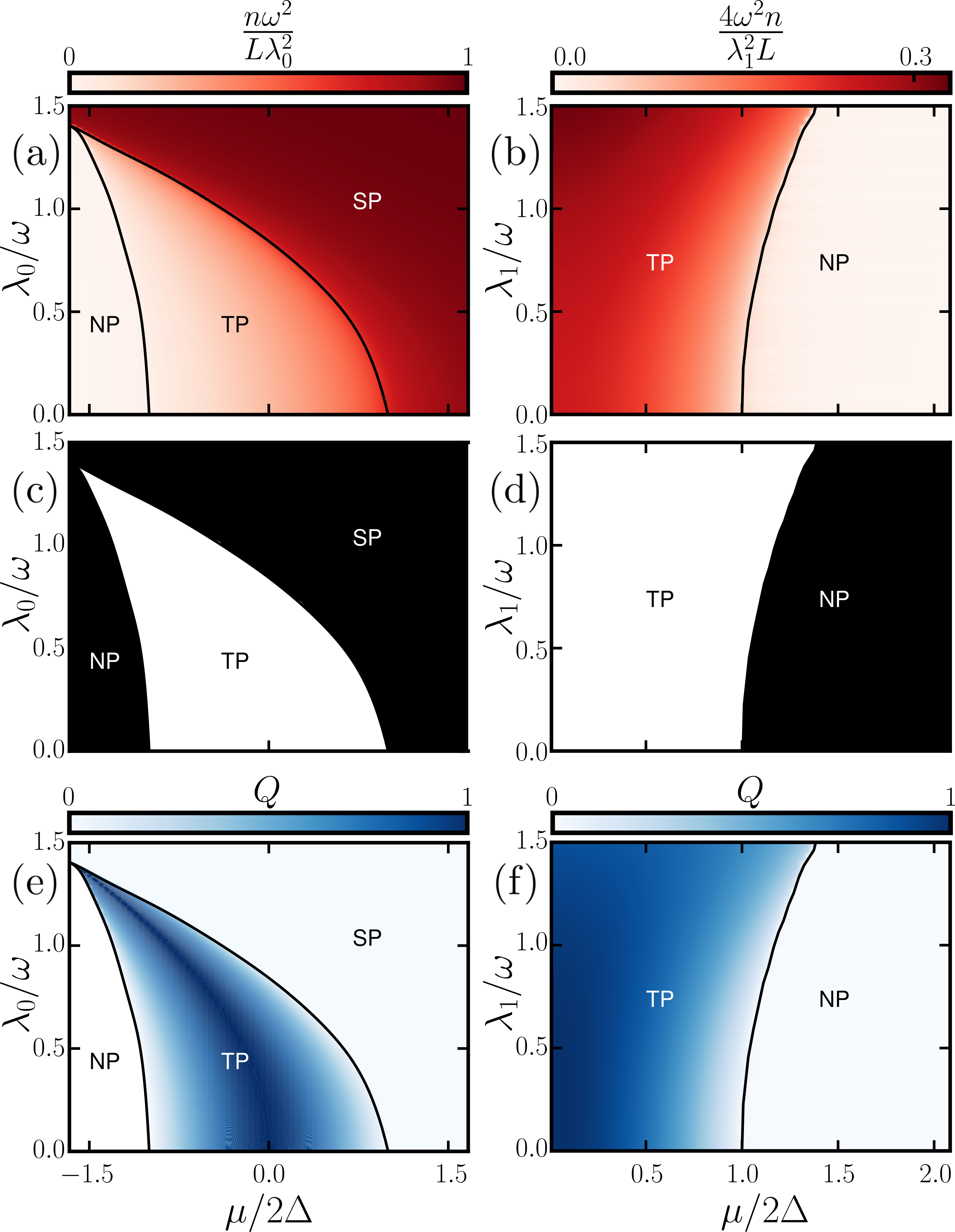}
    \caption{Characterization of the phases of the system. Panels (a), (c) and (e) [(b), (d) and (f)] are results for the chemical potential-like (hoppinglike) coupling. [(a) and (b)] Number of photons normalized by the maximum value predicted by MF in the phase diagrams. [(c) and (d)] Majorana number topological invariant [Eq. \eqref{Eq:MajoranaNumber}] in the phase diagrams. The white regions have $M=-1$ which accounts for the topological phase. The black regions correspond to $M=1$, which represents a phase that is topologically trivial. [(e) and (f)] Two ends correlations $Q$ in the phase diagrams. The Kitaev-cavity parameters are $L=100$ and $\Delta =0.6 \omega$.
    }
    \label{fig:NumberOfPhotons}
    \label{fig:Q_PD}
    \label{fig:MajoranaNumber}
\end{figure}

\section{Phase characterization}
\label{App:PhaseCharacterization}

To understand the many-body phases that are bounded by the critical points, we may recur to different observables of the chain and cavity. Let us start with the state of the cavity. Inspired by the Dicke model, the main observable of the cavity to characterize the phase of the system is the number of photons \cite{DickeModel}. From our analysis we were able to find three different behaviors of radiation in the cavity; namely, trivial, super-radiant and asymptotically super-radiant. All three are displayed in Figs. \ref{fig:NumberOfPhotons}(a) and \ref{fig:NumberOfPhotons}(b) for the two types of light-matter coupling; the borderlines in this figure were signaled by the maximum in the absolute value of the derivative of the number of photons with respect to $\mu$. The trivial behavior of the cavity is characterized by the absence of photons as occurs in the NP for both couplings. We refer to the super-radiant behavior as that where there are photons in the cavity, but their number is lower than the maximum allowed by Eq. \eqref{Minimization}. The coupled system shows a super-radiant when the phase of the chain is topological, that is, when it is located in the TP of the phase diagram. Finally, the asymptotically super-radiant phase is the phase where the number of photons asymptotically reaches the maximum value allowed by the physical restrictions [i.e., Eq. \eqref{Minimization}]. The asymptotically super-radiant phase emerges only with the on-site interaction; the normalized value of the number of photons rapidly approaches 1 in this phase [cf. Fig. \ref{fig:NumberOfPhotons}(a)]. We refer to the latter behavior of the radiance, together with a trivial ordering of the chain, as the SP phase.\\

On the other hand, to describe the phase of the chain we need to assess its topological properties. The Majorana number is a topological invariant that indicates the presence of ZEM in one-dimensional systems \cite{Kitaev}. From our MF approach we can recover the Majorana number by simply replacing the effective chemical potential and hopping in its expression for the isolated Kitaev chain, which yields to
\begin{equation}
\label{Eq:MajoranaNumber}
    M={\rm sgn}(\mu_{\rm eff}-4t_{\rm eff}).
\end{equation}
The Majorana number takes values $=-1$ when a system has a topological ordering and 1 when the ordering is trivial. In Figs. \ref{fig:MajoranaNumber}(c) and \ref{fig:MajoranaNumber}(d) we show the Majorana number behavior for different parameters in the phase diagram; the value of $\mu_{\rm eff}$ and $t_{\rm eff}$ was obtained by DMRG taking the expected value $x$. However, we can recover more information about the state of the chain with the two end correlations $Q$ described in Sec. \ref{Sec:PhaseDiagram}. Since $Q$ is a quantification of the localization of the ZEM, with higher values of $Q$ the emergent Majoranas are expected to be more topologically protected. The behavior of $Q$ for the phase diagram of both couplings is reported in Figs. \ref{fig:Q_PD}(e) and \ref{fig:Q_PD}(f); the boundaries of the phases in this figure were built from the point where $Q$ exceeds $Q_{\rm Trigger}$. The latter was defined as $10^{-2}$ by comparison with exact diagonalization as described in Sec. \ref{Sec:PhaseDiagram}. Nevertheless, the phase transition points obtanied by the Majorana number and the definition of $Q_{\rm Trigger}$ are consistent with each other. Moreover, in Fig. \ref{fig:Q_PD}(e) we can appreciate that the on-site coupling shifts the point of maximum localization of ZEM in the chain. \\

In summary, comparing Figs. \ref{fig:NumberOfPhotons}(a) and \ref{fig:NumberOfPhotons}(b) and Figs. \ref{fig:NumberOfPhotons}(e) and \ref{fig:NumberOfPhotons}(f); we can identify the definition of the phases. The NP is topologically trivial and does not experience radiance in the cavity. The TP is topological in the chain and super-radiant in the cavity. Finally, the SP phase is topologically trivial in the chain and asymptotically super-radiant in the cavity

\section{ Density matrix renormalization group vs mean-field}
\label{App.DMRGVsMF}
\begin{figure}[b!]
\begin{center}
\includegraphics[width=1\linewidth]{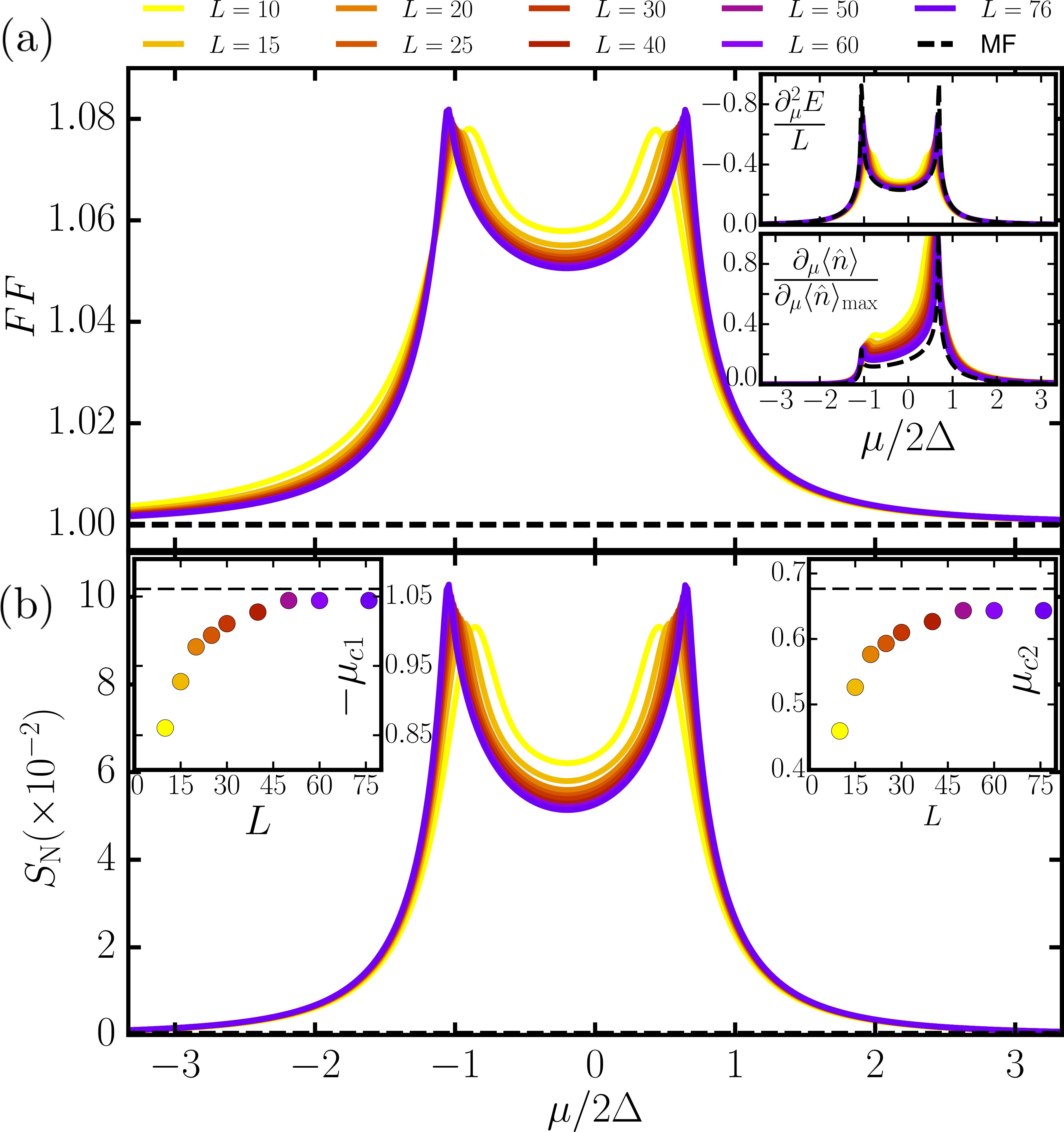}
\end{center}
\caption{Observables for the on-site coupling. (a) FF of the cavity. Insets: Other expected values of the system that show singularities at the critical points of phase transition; each plot depicts the corresponding expected values for different sizes of the chain. The mean-field (MF) results are shown by the dashed black line. Upper inset: Second derivative of the energy with respect to $\mu/2\Delta$. Lower inset: First derivative of the number of photons. Each curve is normalized by the respective maximum. (b) von Neumann entropy of the system with a bipartition between photons and the Kitaev chain. Insets: Scaling of the critical points with the size of the chain; the MF value is shown with the dashed line.  The parameters for these plots are $\omega=1$, $\Delta=0.6\omega$, $\lambda_0=0.49\omega$ and $\lambda_1=0$.}\label{Cavity&Energy}
\end{figure}
For the on-site coupling, we can identify the two second-order phase transitions with the upper inset of Fig.~\ref{Cavity&Energy}(a) for different sizes of the chain; the result shows a rapid convergence to what is obtained from the MF scheme. For an isolated chain, with the parameters used for Fig. \ref{Cavity&Energy}, the phase transition should occur at $\mu/2\Delta=\pm 1$. For a chain-cavity coupled system with $\lambda_0=0.49 \omega$ and $\lambda_1=0\omega$, in contrast, they now occur at $\mu/2\Delta=-1.04\pm 0.02 \text{ and } 0.64 \pm 0.02$. In Fig.~\ref{Cavity&Energy}, we find that the maximum of the FF and the first derivative of the number of photons match with the critical points predicted by the second derivative of the energy. The FF curve shares a similar shape to that of the von Neumann entropy (cf. Fig.~\ref{Cavity&Energy}(b)). This fact, along with the FF behavior, let us conclude that the chain and the cavity increase their entanglement at the critical points, thus driving the cavity into a super Poissonian state.\\
\\
\begin{figure}[t!]
\begin{center}
\includegraphics[width=1\linewidth]{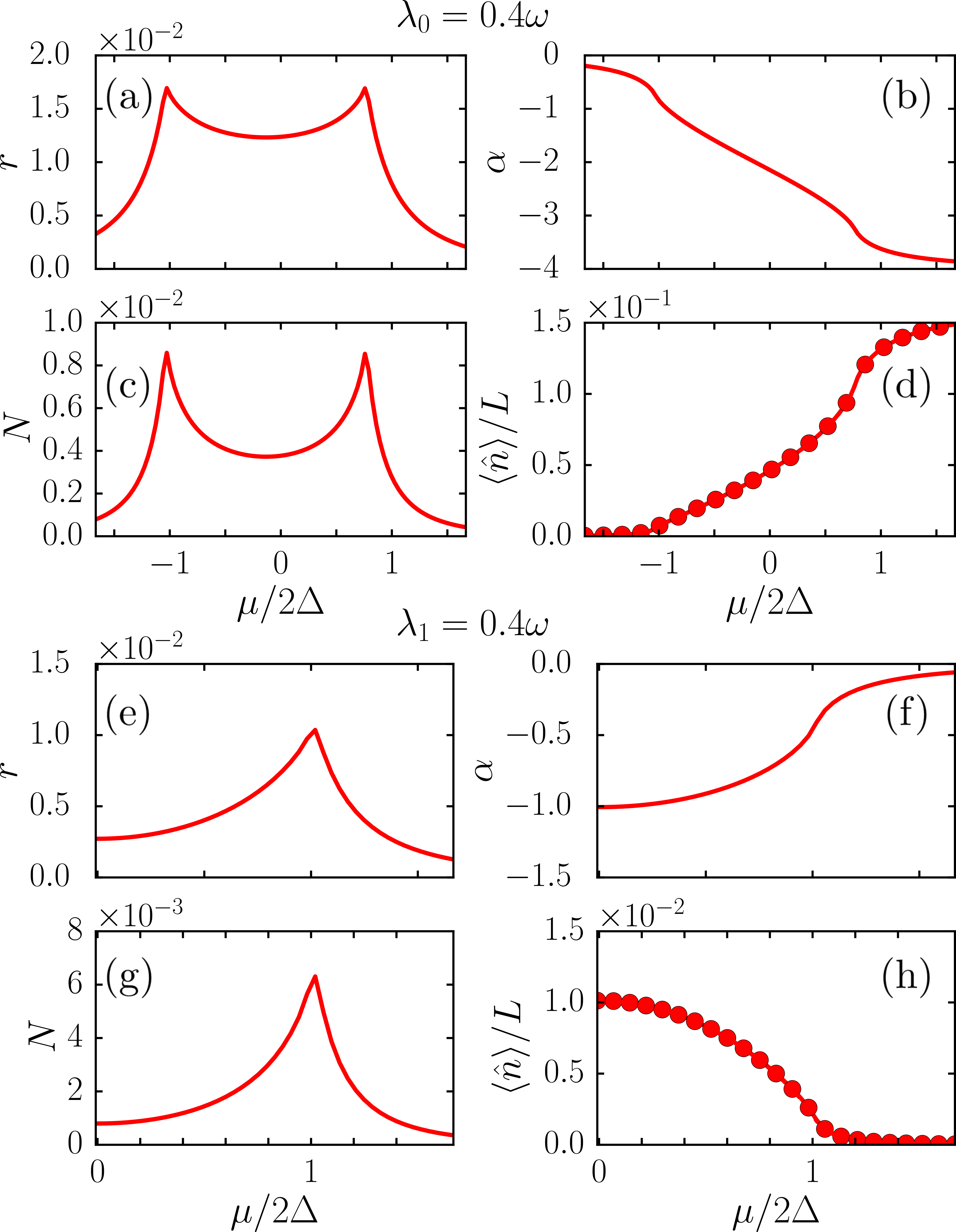}
\end{center}
\caption{Gaussian parameters and number of photons: The squeezing parameter $r$ is depicted in (a) and (e), the coherent parameter $\alpha$ in (b) and (f), the thermal parameter $N$ in (c) and (g). The mean number of photons obtained with DMRG (GS) is shown by symbols (lines) in (d) and (h). The on-site coupling strength is $\lambda_0=0.4\omega$ for (a), (b), (c) and (d). The hoppinglike coupling strength is $\lambda_1=0.4\omega$ for (e), (f), (g) and (h). The other parameters for the cavity-Kitaev system are $L=100$ and $\Delta = 0.6 \omega$.}\label{fig_5:Gaussian}
\end{figure}

The shift of the critical points can be understood considering Eq.~\eqref{eq:FreeEnergy} and the Kitaev chain free energy in Eq.~\eqref{eq:KitaevFreeEnergy}. This last equation, at a fixed $\Delta$ and with $\lambda_0,\text{ }\lambda_1=0$, is a U-shaped curve with a maximum at $\mu=0$, a curve that decays faster as we get farther from the critical point (cf. Fig. \ref{fig:FreeEnergyKitaev}). Then, the free energy of the whole system would allow super-radiance if $\mu_{\rm eff}$ gets farther from the maximum in Eq. \eqref{eq:KitaevFreeEnergy}. Since $\mu_{\rm eff} > \mu$, the system enters slightly earlier into the topological phase (from $-\mu$ to $\mu$) and leaves it highly sooner. The U-shaped free energy will require $x$ to be the minimum possible value as we get farther from the maximum on  the right. As a consequence, in such a region the number of cavity photons will asymptotically approach $n=(\lambda_0/\omega)^2$, generating the SP. Deep into the region at the left of the maximum, the free energy will require the maximum $x$ possible; then $\langle \hat{n}\rangle$ will vanish. In the transition between the nonradiant and asymptotic phases, we can find the TP, which is super-radiant for the cavity. Therefore, the topological phase can be recognized as the phase between peaks in the first derivative of the number of photons, as shown in the lowest inset in Fig. \ref{Cavity&Energy}(a). In the left inset of Fig.~\ref{Cavity&Energy}(b), we can see that the critical points accurately converge to the result predicted by the MF. Slight disagreements are higher for $\mu_{c2}$, since the cavity, and therefore correlations, play a more significant role for that parameter region because it is easier to generate radiation.
\\

%The Eq. \eqref{Minimization} fits the numerical results with differences of about $3$ orders of magnitude less than the observed value. The same holds for the difference between $\langle \hat{n}\rangle$ and $Lx^2$. These differences exhibit peaks at the critical points, but the order of magnitude shows the consistency of the states with MF results. We can observe the site dependence of that relation in Fig. \ref{fig_3:Occupation} with the readings of the occupation number. The value of $x$ for $L=76$ is in excellent agreement with the value obtained for $\langle c_{L/2}^\dagger c_{L/2}\rangle$ (through Eq. (5)), and is even quantitatively accurate for sites close to the edge (see $j=2,4$). The discrepancy within the topological region is high at the edge site, as expected due to the boundary conditions. It is important to note that the global coupling will not provide a direct reading of the chain state at the edge. This is because as the cavity interacts with the whole chain, the expected values of the latter that are extracted from the cavity represent averaged information.\\

For the hoppinglike interaction, the phase transition is symmetrical with respect to $\mu$ since now $\mu_{\rm eff}=\mu$. The effect of $t_{\rm eff}$ in the free energy can be understood again by considering the isolated Kitaev free energy. Regarding $\Delta$ as the independent variable, holding $\mu$ constant and using $\lambda_0,\text{ }\lambda_1=0$, the free energy is again a U-shaped curve with a maximum at $\Delta=0$ (cf. Fig. \ref{fig:FreeEnergyKitaev}). Then, $x$ will allow super-radiance to minimize the free energy of the whole system. The most remarkable result of this kind of nonlocal coupling is the behavior of the cavity as a control device since there are only photons in the cavity when the chain is in the topological phase.
% We can identify an abrupt jump in the number of photons at  $\mu_c/2\Delta= \pm( 1.14\pm0.01)$,  for a chain of $L=76$ sites. Following the $Q$ value, it is at that point where the phase transition occurs, and in the super-radiant region, $-\mu_c<\mu<\mu_c$, we can find the topological phase.

\section{Gaussian Parameters}
\label{App:GaussianParameters}

The fundamental Gaussian information was build from DMRG ground-state expected values as discussed in Secs. \ref{Sec:VNCRiticalityGS}, \ref{Sec:Renyi}. As mentioned in Appendix \ref{App:MFA}, the imaginary part of $\alpha$ is equal to zero, leading to $\alpha\in {\rm I\!R}$ for all parameters; thus $\langle \hat{q}\hat{p} + \hat{p}\hat{q} \rangle =0$. Looking at the results for $N$, $r$ and $\alpha$ the approximations $N, r << \abs{\alpha}$ and  $N, r << 1$ are justified in the analyzed parameter window (cf. Fig. \ref{fig_5:Gaussian}). We also compared the mean number of photons obtained with DMRG and with the Gaussian approximation. The latter leads to
\begin{equation}
    \langle \hat{n}\rangle=(N+1/2)\cosh[2r]+\alpha^2-1/2,
\end{equation}
which under the approximations defined above is $\langle \hat{n}\rangle\approx\alpha ^2 $. As seen in Fig.~\ref{fig_5:Gaussian}, the GS approximation results agree well with those obtained directly from DMRG simulations.

\bibliography{Photon-Fermion}

\begin{thebibliography}{73}%
\makeatletter
\providecommand \@ifxundefined [1]{%
 \@ifx{#1\undefined}
}%
\providecommand \@ifnum [1]{%
 \ifnum #1\expandafter \@firstoftwo
 \else \expandafter \@secondoftwo
 \fi
}%
\providecommand \@ifx [1]{%
 \ifx #1\expandafter \@firstoftwo
 \else \expandafter \@secondoftwo
 \fi
}%
\providecommand \natexlab [1]{#1}%
\providecommand \enquote  [1]{``#1''}%
\providecommand \bibnamefont  [1]{#1}%
\providecommand \bibfnamefont [1]{#1}%
\providecommand \citenamefont [1]{#1}%
\providecommand \href@noop [0]{\@secondoftwo}%
\providecommand \href [0]{\begingroup \@sanitize@url \@href}%
\providecommand \@href[1]{\@@startlink{#1}\@@href}%
\providecommand \@@href[1]{\endgroup#1\@@endlink}%
\providecommand \@sanitize@url [0]{\catcode `\\12\catcode `\$12\catcode
  `\&12\catcode `\#12\catcode `\^12\catcode `\_12\catcode `\%12\relax}%
\providecommand \@@startlink[1]{}%
\providecommand \@@endlink[0]{}%
\providecommand \url  [0]{\begingroup\@sanitize@url \@url }%
\providecommand \@url [1]{\endgroup\@href {#1}{\urlprefix }}%
\providecommand \urlprefix  [0]{URL }%
\providecommand \Eprint [0]{\href }%
\providecommand \doibase [0]{https://doi.org/}%
\providecommand \selectlanguage [0]{\@gobble}%
\providecommand \bibinfo  [0]{\@secondoftwo}%
\providecommand \bibfield  [0]{\@secondoftwo}%
\providecommand \translation [1]{[#1]}%
\providecommand \BibitemOpen [0]{}%
\providecommand \bibitemStop [0]{}%
\providecommand \bibitemNoStop [0]{.\EOS\space}%
\providecommand \EOS [0]{\spacefactor3000\relax}%
\providecommand \BibitemShut  [1]{\csname bibitem#1\endcsname}%
\let\auto@bib@innerbib\@empty
%</preamble>
\bibitem [{\citenamefont {Rossini}\ and\ \citenamefont
  {Fazio}(2012)}]{ExtendedBH}%
  \BibitemOpen
  \bibfield  {author} {\bibinfo {author} {\bibfnamefont {D.}~\bibnamefont
  {Rossini}}\ and\ \bibinfo {author} {\bibfnamefont {R.}~\bibnamefont
  {Fazio}},\ }\bibfield  {title} {\bibinfo {title} {Phase diagram of the
  extended {B}ose-{H}ubbard model},\ }\href
  {http://stacks.iop.org/1367-2630/14/i=6/a=065012} {\bibfield  {journal}
  {\bibinfo  {journal} {New J. Phys.}\ }\textbf {\bibinfo {volume} {14}},\
  \bibinfo {pages} {065012} (\bibinfo {year} {2012})}\BibitemShut {NoStop}%
\bibitem [{\citenamefont {Lewis-Swan}\ \emph {et~al.}(2019)\citenamefont
  {Lewis-Swan}, \citenamefont {Safavi-Naini}, \citenamefont {Bollinger},\ and\
  \citenamefont {Rey}}]{Lewis_SwanDM}%
  \BibitemOpen
  \bibfield  {author} {\bibinfo {author} {\bibfnamefont {R.~J.}\ \bibnamefont
  {Lewis-Swan}}, \bibinfo {author} {\bibfnamefont {A.}~\bibnamefont
  {Safavi-Naini}}, \bibinfo {author} {\bibfnamefont {J.~J.}\ \bibnamefont
  {Bollinger}},\ and\ \bibinfo {author} {\bibfnamefont {A.~M.}\ \bibnamefont
  {Rey}},\ }\bibfield  {title} {\bibinfo {title} {Unifying scrambling,
  thermalization and entanglement through measurement of fidelity
  out-of-time-order correlators in the {D}icke model},\ }\href
  {http://dx.doi.org/10.1038/s41467-019-09436-y} {\bibfield  {journal}
  {\bibinfo  {journal} {Nat. Commun.}\ }\textbf {\bibinfo {volume} {10}},\
  \bibinfo {pages} {1581} (\bibinfo {year} {2019})}\BibitemShut {NoStop}%
\bibitem [{\citenamefont {Elben}\ \emph {et~al.}(2018)\citenamefont {Elben},
  \citenamefont {Vermersch}, \citenamefont {Dalmonte}, \citenamefont {Cirac},\
  and\ \citenamefont {Zoller}}]{REHubbard&Spin}%
  \BibitemOpen
  \bibfield  {author} {\bibinfo {author} {\bibfnamefont {A.}~\bibnamefont
  {Elben}}, \bibinfo {author} {\bibfnamefont {B.}~\bibnamefont {Vermersch}},
  \bibinfo {author} {\bibfnamefont {M.}~\bibnamefont {Dalmonte}}, \bibinfo
  {author} {\bibfnamefont {J.~I.}\ \bibnamefont {Cirac}},\ and\ \bibinfo
  {author} {\bibfnamefont {P.}~\bibnamefont {Zoller}},\ }\bibfield  {title}
  {\bibinfo {title} {{R}\'enyi entropies from random quenches in atomic
  {H}ubbard and spin models},\ }\href
  {https://doi.org/10.1103/PhysRevLett.120.050406} {\bibfield  {journal}
  {\bibinfo  {journal} {Phys. Rev. Lett.}\ }\textbf {\bibinfo {volume} {120}},\
  \bibinfo {pages} {050406} (\bibinfo {year} {2018})}\BibitemShut {NoStop}%
\bibitem [{\citenamefont {Brydges}\ \emph {et~al.}(2019)\citenamefont
  {Brydges}, \citenamefont {Elben}, \citenamefont {Jurcevic}, \citenamefont
  {Vermersch}, \citenamefont {Maier}, \citenamefont {Lanyon}, \citenamefont
  {Zoller}, \citenamefont {Blatt},\ and\ \citenamefont
  {Roos}}]{BrydgesRERandomized}%
  \BibitemOpen
  \bibfield  {author} {\bibinfo {author} {\bibfnamefont {T.}~\bibnamefont
  {Brydges}}, \bibinfo {author} {\bibfnamefont {A.}~\bibnamefont {Elben}},
  \bibinfo {author} {\bibfnamefont {P.}~\bibnamefont {Jurcevic}}, \bibinfo
  {author} {\bibfnamefont {B.}~\bibnamefont {Vermersch}}, \bibinfo {author}
  {\bibfnamefont {C.}~\bibnamefont {Maier}}, \bibinfo {author} {\bibfnamefont
  {B.~P.}\ \bibnamefont {Lanyon}}, \bibinfo {author} {\bibfnamefont
  {P.}~\bibnamefont {Zoller}}, \bibinfo {author} {\bibfnamefont
  {R.}~\bibnamefont {Blatt}},\ and\ \bibinfo {author} {\bibfnamefont {C.~F.}\
  \bibnamefont {Roos}},\ }\bibfield  {title} {\bibinfo {title} {Probing
  {R}ényi entanglement entropy via randomized measurements},\ }\href
  {https://doi.org/10.1126/science.aau4963} {\bibfield  {journal} {\bibinfo
  {journal} {Science}\ }\textbf {\bibinfo {volume} {364}},\ \bibinfo {pages}
  {260–263} (\bibinfo {year} {2019})}\BibitemShut {NoStop}%
\bibitem [{\citenamefont {Camacho-Guardian}\ \emph {et~al.}(2017)\citenamefont
  {Camacho-Guardian}, \citenamefont {Paredes},\ and\ \citenamefont
  {Caballero-Ben\'{\i}tez}}]{Camacho_Guardian_2017OpticalLattices}%
  \BibitemOpen
  \bibfield  {author} {\bibinfo {author} {\bibfnamefont {A.}~\bibnamefont
  {Camacho-Guardian}}, \bibinfo {author} {\bibfnamefont {R.}~\bibnamefont
  {Paredes}},\ and\ \bibinfo {author} {\bibfnamefont {S.~F.}\ \bibnamefont
  {Caballero-Ben\'{\i}tez}},\ }\bibfield  {title} {\bibinfo {title} {Quantum
  simulation of competing orders with fermions in quantum optical lattices},\
  }\href {https://doi.org/10.1103/PhysRevA.96.051602} {\bibfield  {journal}
  {\bibinfo  {journal} {Phys. Rev. A}\ }\textbf {\bibinfo {volume} {96}},\
  \bibinfo {pages} {051602(R)} (\bibinfo {year} {2017})}\BibitemShut {NoStop}%
\bibitem [{\citenamefont {Zhang}\ \emph
  {et~al.}(2018{\natexlab{a}})\citenamefont {Zhang}, \citenamefont {Zhang},
  \citenamefont {Shen}, \citenamefont {Zhang}, \citenamefont {Zhang},
  \citenamefont {Yung}, \citenamefont {Casanova}, \citenamefont {Pedernales},
  \citenamefont {Lamata}, \citenamefont {Solano},\ and\ \citenamefont
  {Kim}}]{ExpFERMBOSONS}%
  \BibitemOpen
  \bibfield  {author} {\bibinfo {author} {\bibfnamefont {X.}~\bibnamefont
  {Zhang}}, \bibinfo {author} {\bibfnamefont {K.}~\bibnamefont {Zhang}},
  \bibinfo {author} {\bibfnamefont {Y.}~\bibnamefont {Shen}}, \bibinfo {author}
  {\bibfnamefont {S.}~\bibnamefont {Zhang}}, \bibinfo {author} {\bibfnamefont
  {J.-N.}\ \bibnamefont {Zhang}}, \bibinfo {author} {\bibfnamefont {M.-H.}\
  \bibnamefont {Yung}}, \bibinfo {author} {\bibfnamefont {J.}~\bibnamefont
  {Casanova}}, \bibinfo {author} {\bibfnamefont {J.~S.}\ \bibnamefont
  {Pedernales}}, \bibinfo {author} {\bibfnamefont {L.}~\bibnamefont {Lamata}},
  \bibinfo {author} {\bibfnamefont {E.}~\bibnamefont {Solano}},\ and\ \bibinfo
  {author} {\bibfnamefont {K.}~\bibnamefont {Kim}},\ }\bibfield  {title}
  {\bibinfo {title} {Experimental quantum simulation of fermion-antifermion
  scattering via boson exchange in a trapped ion},\ }\href
  {https://doi.org/10.1038/s41467-017-02507-y} {\bibfield  {journal} {\bibinfo
  {journal} {Nat. Commun.}\ }\textbf {\bibinfo {volume} {9}},\ \bibinfo {pages}
  {195} (\bibinfo {year} {2018}{\natexlab{a}})}\BibitemShut {NoStop}%
\bibitem [{\citenamefont {Baumann}\ \emph {et~al.}(2010)\citenamefont
  {Baumann}, \citenamefont {Guerlin}, \citenamefont {Brennecke},\ and\
  \citenamefont {Esslinger}}]{ExpDickeSuperfluid}%
  \BibitemOpen
  \bibfield  {author} {\bibinfo {author} {\bibfnamefont {K.}~\bibnamefont
  {Baumann}}, \bibinfo {author} {\bibfnamefont {C.}~\bibnamefont {Guerlin}},
  \bibinfo {author} {\bibfnamefont {F.}~\bibnamefont {Brennecke}},\ and\
  \bibinfo {author} {\bibfnamefont {T.}~\bibnamefont {Esslinger}},\ }\bibfield
  {title} {\bibinfo {title} {Dicke quantum phase transition with a superfluid
  gas in an optical cavity},\ }\href {https://doi.org/10.1038/nature09009}
  {\bibfield  {journal} {\bibinfo  {journal} {Nature}\ }\textbf {\bibinfo
  {volume} {464}},\ \bibinfo {pages} {1301} (\bibinfo {year}
  {2010})}\BibitemShut {NoStop}%
\bibitem [{\citenamefont {Baumann}\ \emph {et~al.}(2011)\citenamefont
  {Baumann}, \citenamefont {Mottl}, \citenamefont {Brennecke},\ and\
  \citenamefont {Esslinger}}]{ExpDickeSuperfluidPRL}%
  \BibitemOpen
  \bibfield  {author} {\bibinfo {author} {\bibfnamefont {K.}~\bibnamefont
  {Baumann}}, \bibinfo {author} {\bibfnamefont {R.}~\bibnamefont {Mottl}},
  \bibinfo {author} {\bibfnamefont {F.}~\bibnamefont {Brennecke}},\ and\
  \bibinfo {author} {\bibfnamefont {T.}~\bibnamefont {Esslinger}},\ }\bibfield
  {title} {\bibinfo {title} {Exploring symmetry breaking at the {D}icke quantum
  phase transition},\ }\href {https://doi.org/10.1103/PhysRevLett.107.140402}
  {\bibfield  {journal} {\bibinfo  {journal} {Phys. Rev. Lett.}\ }\textbf
  {\bibinfo {volume} {107}},\ \bibinfo {pages} {140402} (\bibinfo {year}
  {2011})}\BibitemShut {NoStop}%
\bibitem [{\citenamefont {Léonard}\ \emph {et~al.}(2017)\citenamefont
  {Léonard}, \citenamefont {Morales}, \citenamefont {Zupancic}, \citenamefont
  {Esslinger},\ and\ \citenamefont {Donner}}]{L_onard_2017BECCAvity}%
  \BibitemOpen
  \bibfield  {author} {\bibinfo {author} {\bibfnamefont {J.}~\bibnamefont
  {Léonard}}, \bibinfo {author} {\bibfnamefont {A.}~\bibnamefont {Morales}},
  \bibinfo {author} {\bibfnamefont {P.}~\bibnamefont {Zupancic}}, \bibinfo
  {author} {\bibfnamefont {T.}~\bibnamefont {Esslinger}},\ and\ \bibinfo
  {author} {\bibfnamefont {T.}~\bibnamefont {Donner}},\ }\bibfield  {title}
  {\bibinfo {title} {Supersolid formation in a quantum gas breaking a
  continuous translational symmetry},\ }\href
  {http://dx.doi.org/10.1038/nature21067} {\bibfield  {journal} {\bibinfo
  {journal} {Nature}\ }\textbf {\bibinfo {volume} {543}},\ \bibinfo {pages}
  {87–90} (\bibinfo {year} {2017})}\BibitemShut {NoStop}%
\bibitem [{\citenamefont {Roux}\ \emph {et~al.}(2020)\citenamefont {Roux},
  \citenamefont {Konishi}, \citenamefont {Helson},\ and\ \citenamefont
  {Brantut}}]{roux2020nat}%
  \BibitemOpen
  \bibfield  {author} {\bibinfo {author} {\bibfnamefont {K.}~\bibnamefont
  {Roux}}, \bibinfo {author} {\bibfnamefont {H.}~\bibnamefont {Konishi}},
  \bibinfo {author} {\bibfnamefont {V.}~\bibnamefont {Helson}},\ and\ \bibinfo
  {author} {\bibfnamefont {J.-P.}\ \bibnamefont {Brantut}},\ }\bibfield
  {title} {\bibinfo {title} {{Strongly correlated Fermions strongly coupled to
  light}},\ }\href {https://doi.org/10.1038/s41467-020-16767-8} {\bibfield
  {journal} {\bibinfo  {journal} {Nat. Commun.}\ }\textbf {\bibinfo {volume}
  {11}},\ \bibinfo {pages} {2974} (\bibinfo {year} {2020})}\BibitemShut
  {NoStop}%
\bibitem [{\citenamefont {Kiffner}\ \emph {et~al.}(2019)\citenamefont
  {Kiffner}, \citenamefont {Coulthard}, \citenamefont {Schlawin}, \citenamefont
  {Ardavan},\ and\ \citenamefont {Jaksch}}]{kiffner2019prb}%
  \BibitemOpen
  \bibfield  {author} {\bibinfo {author} {\bibfnamefont {M.}~\bibnamefont
  {Kiffner}}, \bibinfo {author} {\bibfnamefont {J.~R.}\ \bibnamefont
  {Coulthard}}, \bibinfo {author} {\bibfnamefont {F.}~\bibnamefont {Schlawin}},
  \bibinfo {author} {\bibfnamefont {A.}~\bibnamefont {Ardavan}},\ and\ \bibinfo
  {author} {\bibfnamefont {D.}~\bibnamefont {Jaksch}},\ }\bibfield  {title}
  {\bibinfo {title} {Manipulating quantum materials with quantum light},\
  }\href {https://doi.org/10.1103/PhysRevB.99.085116} {\bibfield  {journal}
  {\bibinfo  {journal} {Phys. Rev. B}\ }\textbf {\bibinfo {volume} {99}},\
  \bibinfo {pages} {085116} (\bibinfo {year} {2019})}\BibitemShut {NoStop}%
\bibitem [{\citenamefont {Thomas}\ \emph {et~al.}()\citenamefont {Thomas},
  \citenamefont {Devaux}, \citenamefont {Nagarajan}, \citenamefont {Chervy},
  \citenamefont {Seidel}, \citenamefont {Hagenmüller}, \citenamefont
  {Schütz}, \citenamefont {Schachenmayer}, \citenamefont {Genet},
  \citenamefont {Pupillo},\ and\ \citenamefont {Ebbesen}}]{thomasEFSC}%
  \BibitemOpen
  \bibfield  {author} {\bibinfo {author} {\bibfnamefont {A.}~\bibnamefont
  {Thomas}}, \bibinfo {author} {\bibfnamefont {E.}~\bibnamefont {Devaux}},
  \bibinfo {author} {\bibfnamefont {K.}~\bibnamefont {Nagarajan}}, \bibinfo
  {author} {\bibfnamefont {T.}~\bibnamefont {Chervy}}, \bibinfo {author}
  {\bibfnamefont {M.}~\bibnamefont {Seidel}}, \bibinfo {author} {\bibfnamefont
  {D.}~\bibnamefont {Hagenmüller}}, \bibinfo {author} {\bibfnamefont
  {S.}~\bibnamefont {Schütz}}, \bibinfo {author} {\bibfnamefont
  {J.}~\bibnamefont {Schachenmayer}}, \bibinfo {author} {\bibfnamefont
  {C.}~\bibnamefont {Genet}}, \bibinfo {author} {\bibfnamefont
  {G.}~\bibnamefont {Pupillo}},\ and\ \bibinfo {author} {\bibfnamefont {T.~W.}\
  \bibnamefont {Ebbesen}},\ }\href@noop {} {}\Eprint
  {https://arxiv.org/abs/1911.01459} {arXiv:1911.01459} \BibitemShut {NoStop}%
\bibitem [{\citenamefont {Curtis}\ \emph {et~al.}(2019)\citenamefont {Curtis},
  \citenamefont {Raines}, \citenamefont {Allocca}, \citenamefont {Hafezi},\
  and\ \citenamefont {Galitski}}]{curtis2019prl}%
  \BibitemOpen
  \bibfield  {author} {\bibinfo {author} {\bibfnamefont {J.~B.}\ \bibnamefont
  {Curtis}}, \bibinfo {author} {\bibfnamefont {Z.~M.}\ \bibnamefont {Raines}},
  \bibinfo {author} {\bibfnamefont {A.~A.}\ \bibnamefont {Allocca}}, \bibinfo
  {author} {\bibfnamefont {M.}~\bibnamefont {Hafezi}},\ and\ \bibinfo {author}
  {\bibfnamefont {V.~M.}\ \bibnamefont {Galitski}},\ }\bibfield  {title}
  {\bibinfo {title} {{Cavity Quantum Eliashberg Enhancement of
  Superconductivity}},\ }\href {https://doi.org/10.1103/PhysRevLett.122.167002}
  {\bibfield  {journal} {\bibinfo  {journal} {Phys. Rev. Lett.}\ }\textbf
  {\bibinfo {volume} {122}},\ \bibinfo {pages} {167002} (\bibinfo {year}
  {2019})}\BibitemShut {NoStop}%
\bibitem [{\citenamefont {Schlawin}\ \emph {et~al.}(2019)\citenamefont
  {Schlawin}, \citenamefont {Cavalleri},\ and\ \citenamefont
  {Jaksch}}]{FrankEPSC}%
  \BibitemOpen
  \bibfield  {author} {\bibinfo {author} {\bibfnamefont {F.}~\bibnamefont
  {Schlawin}}, \bibinfo {author} {\bibfnamefont {A.}~\bibnamefont
  {Cavalleri}},\ and\ \bibinfo {author} {\bibfnamefont {D.}~\bibnamefont
  {Jaksch}},\ }\bibfield  {title} {\bibinfo {title} {Cavity-mediated
  electron-photon superconductivity},\ }\href
  {https://doi.org/10.1103/PhysRevLett.122.133602} {\bibfield  {journal}
  {\bibinfo  {journal} {Phys. Rev. Lett.}\ }\textbf {\bibinfo {volume} {122}},\
  \bibinfo {pages} {133602} (\bibinfo {year} {2019})}\BibitemShut {NoStop}%
\bibitem [{\citenamefont {Gao}\ \emph {et~al.}(2020)\citenamefont {Gao},
  \citenamefont {Schlawin}, \citenamefont {Buzzi}, \citenamefont {Cavalleri},\
  and\ \citenamefont {Jaksch}}]{GaoPEP}%
  \BibitemOpen
  \bibfield  {author} {\bibinfo {author} {\bibfnamefont {H.}~\bibnamefont
  {Gao}}, \bibinfo {author} {\bibfnamefont {F.}~\bibnamefont {Schlawin}},
  \bibinfo {author} {\bibfnamefont {M.}~\bibnamefont {Buzzi}}, \bibinfo
  {author} {\bibfnamefont {A.}~\bibnamefont {Cavalleri}},\ and\ \bibinfo
  {author} {\bibfnamefont {D.}~\bibnamefont {Jaksch}},\ }\bibfield  {title}
  {\bibinfo {title} {Photoinduced electron pairing in a driven cavity},\ }\href
  {https://doi.org/10.1103/PhysRevLett.125.053602} {\bibfield  {journal}
  {\bibinfo  {journal} {Phys. Rev. Lett.}\ }\textbf {\bibinfo {volume} {125}},\
  \bibinfo {pages} {053602} (\bibinfo {year} {2020})}\BibitemShut {NoStop}%
\bibitem [{\citenamefont {Forn-D\'{\i}az}\ \emph {et~al.}(2019)\citenamefont
  {Forn-D\'{\i}az}, \citenamefont {Lamata}, \citenamefont {Rico}, \citenamefont
  {Kono},\ and\ \citenamefont {Solano}}]{FornRMP2019}%
  \BibitemOpen
  \bibfield  {author} {\bibinfo {author} {\bibfnamefont {P.}~\bibnamefont
  {Forn-D\'{\i}az}}, \bibinfo {author} {\bibfnamefont {L.}~\bibnamefont
  {Lamata}}, \bibinfo {author} {\bibfnamefont {E.}~\bibnamefont {Rico}},
  \bibinfo {author} {\bibfnamefont {J.}~\bibnamefont {Kono}},\ and\ \bibinfo
  {author} {\bibfnamefont {E.}~\bibnamefont {Solano}},\ }\bibfield  {title}
  {\bibinfo {title} {Ultrastrong coupling regimes of light-matter
  interaction},\ }\href {https://doi.org/10.1103/RevModPhys.91.025005}
  {\bibfield  {journal} {\bibinfo  {journal} {Rev. Mod. Phys.}\ }\textbf
  {\bibinfo {volume} {91}},\ \bibinfo {pages} {025005} (\bibinfo {year}
  {2019})}\BibitemShut {NoStop}%
\bibitem [{\citenamefont {Frisk~Kockum}\ \emph {et~al.}(2019)\citenamefont
  {Frisk~Kockum}, \citenamefont {Miranowicz}, \citenamefont {De~Liberato},
  \citenamefont {Savasta},\ and\ \citenamefont {Nori}}]{FriskNPR2019}%
  \BibitemOpen
  \bibfield  {author} {\bibinfo {author} {\bibfnamefont {A.}~\bibnamefont
  {Frisk~Kockum}}, \bibinfo {author} {\bibfnamefont {A.}~\bibnamefont
  {Miranowicz}}, \bibinfo {author} {\bibfnamefont {S.}~\bibnamefont
  {De~Liberato}}, \bibinfo {author} {\bibfnamefont {S.}~\bibnamefont
  {Savasta}},\ and\ \bibinfo {author} {\bibfnamefont {F.}~\bibnamefont
  {Nori}},\ }\bibfield  {title} {\bibinfo {title} {Ultrastrong coupling between
  light and matter},\ }\href {https://doi.org/10.1038/s42254-018-0006-2}
  {\bibfield  {journal} {\bibinfo  {journal} {Nat. Rev. Phys.}\ }\textbf
  {\bibinfo {volume} {1}},\ \bibinfo {pages} {19} (\bibinfo {year}
  {2019})}\BibitemShut {NoStop}%
\bibitem [{\citenamefont {Dartiailh}\ \emph {et~al.}(2017)\citenamefont
  {Dartiailh}, \citenamefont {Kontos}, \citenamefont {Dou\c{c}ot},\ and\
  \citenamefont {Cottet}}]{DartiailhMajoranaPairs}%
  \BibitemOpen
  \bibfield  {author} {\bibinfo {author} {\bibfnamefont {M.~C.}\ \bibnamefont
  {Dartiailh}}, \bibinfo {author} {\bibfnamefont {T.}~\bibnamefont {Kontos}},
  \bibinfo {author} {\bibfnamefont {B.}~\bibnamefont {Dou\c{c}ot}},\ and\
  \bibinfo {author} {\bibfnamefont {A.}~\bibnamefont {Cottet}},\ }\bibfield
  {title} {\bibinfo {title} {Direct cavity detection of {M}ajorana pairs},\
  }\href {https://doi.org/10.1103/PhysRevLett.118.126803} {\bibfield  {journal}
  {\bibinfo  {journal} {Phys. Rev. Lett.}\ }\textbf {\bibinfo {volume} {118}},\
  \bibinfo {pages} {126803} (\bibinfo {year} {2017})}\BibitemShut {NoStop}%
\bibitem [{\citenamefont {Schlawin}\ and\ \citenamefont
  {Jaksch}(2019)}]{frank2019prl}%
  \BibitemOpen
  \bibfield  {author} {\bibinfo {author} {\bibfnamefont {F.}~\bibnamefont
  {Schlawin}}\ and\ \bibinfo {author} {\bibfnamefont {D.}~\bibnamefont
  {Jaksch}},\ }\bibfield  {title} {\bibinfo {title} {Cavity-mediated
  unconventional pairing in ultracold fermionic atoms},\ }\href
  {https://doi.org/10.1103/PhysRevLett.123.133601} {\bibfield  {journal}
  {\bibinfo  {journal} {Phys. Rev. Lett.}\ }\textbf {\bibinfo {volume} {123}},\
  \bibinfo {pages} {133601} (\bibinfo {year} {2019})}\BibitemShut {NoStop}%
\bibitem [{\citenamefont {Nie}\ \emph {et~al.}(2020)\citenamefont {Nie},
  \citenamefont {Peng}, \citenamefont {Nori},\ and\ \citenamefont
  {Liu}}]{Nie_2020Superatom}%
  \BibitemOpen
  \bibfield  {author} {\bibinfo {author} {\bibfnamefont {W.}~\bibnamefont
  {Nie}}, \bibinfo {author} {\bibfnamefont {Z.~H.}\ \bibnamefont {Peng}},
  \bibinfo {author} {\bibfnamefont {F.}~\bibnamefont {Nori}},\ and\ \bibinfo
  {author} {\bibfnamefont {Y.-x.}\ \bibnamefont {Liu}},\ }\bibfield  {title}
  {\bibinfo {title} {Topologically protected quantum coherence in a
  superatom},\ }\href {https://doi.org/10.1103/PhysRevLett.124.023603}
  {\bibfield  {journal} {\bibinfo  {journal} {Phys. Rev. Lett.}\ }\textbf
  {\bibinfo {volume} {124}},\ \bibinfo {pages} {023603} (\bibinfo {year}
  {2020})}\BibitemShut {NoStop}%
\bibitem [{\citenamefont {Mourik}\ \emph {et~al.}(2012)\citenamefont {Mourik},
  \citenamefont {Zuo}, \citenamefont {Frolov}, \citenamefont {Plissard},
  \citenamefont {Bakkers},\ and\ \citenamefont {Kouwenhoven}}]{Mourik1003}%
  \BibitemOpen
  \bibfield  {author} {\bibinfo {author} {\bibfnamefont {V.}~\bibnamefont
  {Mourik}}, \bibinfo {author} {\bibfnamefont {K.}~\bibnamefont {Zuo}},
  \bibinfo {author} {\bibfnamefont {S.~M.}\ \bibnamefont {Frolov}}, \bibinfo
  {author} {\bibfnamefont {S.~R.}\ \bibnamefont {Plissard}}, \bibinfo {author}
  {\bibfnamefont {E.~P. A.~M.}\ \bibnamefont {Bakkers}},\ and\ \bibinfo
  {author} {\bibfnamefont {L.~P.}\ \bibnamefont {Kouwenhoven}},\ }\bibfield
  {title} {\bibinfo {title} {Signatures of {M}ajorana fermions in hybrid
  superconductor-semiconductor nanowire devices},\ }\href
  {https://doi.org/10.1126/science.1222360} {\bibfield  {journal} {\bibinfo
  {journal} {Science}\ }\textbf {\bibinfo {volume} {336}},\ \bibinfo {pages}
  {1003} (\bibinfo {year} {2012})}\BibitemShut {NoStop}%
\bibitem [{\citenamefont {Nadj-Perge}\ \emph {et~al.}(2014)\citenamefont
  {Nadj-Perge}, \citenamefont {Drozdov}, \citenamefont {Li}, \citenamefont
  {Chen}, \citenamefont {Jeon}, \citenamefont {Seo}, \citenamefont {MacDonald},
  \citenamefont {Bernevig},\ and\ \citenamefont {Yazdani}}]{ExpMajorana}%
  \BibitemOpen
  \bibfield  {author} {\bibinfo {author} {\bibfnamefont {S.}~\bibnamefont
  {Nadj-Perge}}, \bibinfo {author} {\bibfnamefont {I.~K.}\ \bibnamefont
  {Drozdov}}, \bibinfo {author} {\bibfnamefont {J.}~\bibnamefont {Li}},
  \bibinfo {author} {\bibfnamefont {H.}~\bibnamefont {Chen}}, \bibinfo {author}
  {\bibfnamefont {S.}~\bibnamefont {Jeon}}, \bibinfo {author} {\bibfnamefont
  {J.}~\bibnamefont {Seo}}, \bibinfo {author} {\bibfnamefont {A.~H.}\
  \bibnamefont {MacDonald}}, \bibinfo {author} {\bibfnamefont {B.~A.}\
  \bibnamefont {Bernevig}},\ and\ \bibinfo {author} {\bibfnamefont
  {A.}~\bibnamefont {Yazdani}},\ }\bibfield  {title} {\bibinfo {title}
  {Observation of {M}ajorana fermions in ferromagnetic atomic chains on a
  superconductor},\ }\href
  {https://science.sciencemag.org/content/346/6209/602} {\bibfield  {journal}
  {\bibinfo  {journal} {Science}\ }\textbf {\bibinfo {volume} {346}},\ \bibinfo
  {pages} {602} (\bibinfo {year} {2014})}\BibitemShut {NoStop}%
\bibitem [{\citenamefont {Albrecht}\ \emph {et~al.}(2016)\citenamefont
  {Albrecht}, \citenamefont {Higginbotham}, \citenamefont {Madsen},
  \citenamefont {Kuemmeth}, \citenamefont {Jespersen}, \citenamefont {Nygård},
  \citenamefont {Krogstrup},\ and\ \citenamefont {Marcus}}]{Exp2016}%
  \BibitemOpen
  \bibfield  {author} {\bibinfo {author} {\bibfnamefont {S.~M.}\ \bibnamefont
  {Albrecht}}, \bibinfo {author} {\bibfnamefont {A.~P.}\ \bibnamefont
  {Higginbotham}}, \bibinfo {author} {\bibfnamefont {M.}~\bibnamefont
  {Madsen}}, \bibinfo {author} {\bibfnamefont {F.}~\bibnamefont {Kuemmeth}},
  \bibinfo {author} {\bibfnamefont {T.~S.}\ \bibnamefont {Jespersen}}, \bibinfo
  {author} {\bibfnamefont {J.}~\bibnamefont {Nygård}}, \bibinfo {author}
  {\bibfnamefont {P.}~\bibnamefont {Krogstrup}},\ and\ \bibinfo {author}
  {\bibfnamefont {C.~M.}\ \bibnamefont {Marcus}},\ }\bibfield  {title}
  {\bibinfo {title} {Exponential protection of zero modes in {M}ajorana
  islands},\ }\href {https://doi.org/10.1038/nature17162} {\bibfield  {journal}
  {\bibinfo  {journal} {Nature}\ }\textbf {\bibinfo {volume} {531}},\ \bibinfo
  {pages} {206–209} (\bibinfo {year} {2016})}\BibitemShut {NoStop}%
\bibitem [{\citenamefont {Zhang}\ \emph
  {et~al.}(2018{\natexlab{b}})\citenamefont {Zhang}, \citenamefont {Liu},
  \citenamefont {Gazibegovic}, \citenamefont {Xu}, \citenamefont {Logan},
  \citenamefont {Wang}, \citenamefont {van Loo}, \citenamefont {Bommer},
  \citenamefont {de~Moor}, \citenamefont {Car},\ and\ \citenamefont
  {et~al.}}]{Exp2018}%
  \BibitemOpen
  \bibfield  {author} {\bibinfo {author} {\bibfnamefont {H.}~\bibnamefont
  {Zhang}}, \bibinfo {author} {\bibfnamefont {C.-X.}\ \bibnamefont {Liu}},
  \bibinfo {author} {\bibfnamefont {S.}~\bibnamefont {Gazibegovic}}, \bibinfo
  {author} {\bibfnamefont {D.}~\bibnamefont {Xu}}, \bibinfo {author}
  {\bibfnamefont {J.~A.}\ \bibnamefont {Logan}}, \bibinfo {author}
  {\bibfnamefont {G.}~\bibnamefont {Wang}}, \bibinfo {author} {\bibfnamefont
  {N.}~\bibnamefont {van Loo}}, \bibinfo {author} {\bibfnamefont {J.~D.~S.}\
  \bibnamefont {Bommer}}, \bibinfo {author} {\bibfnamefont {M.~W.~A.}\
  \bibnamefont {de~Moor}}, \bibinfo {author} {\bibfnamefont {D.}~\bibnamefont
  {Car}},\ and\ \bibinfo {author} {\bibnamefont {et~al.}},\ }\bibfield  {title}
  {\bibinfo {title} {Quantized {M}ajorana conductance},\ }\href
  {https://doi.org/10.1038/nature26142} {\bibfield  {journal} {\bibinfo
  {journal} {Nature}\ }\textbf {\bibinfo {volume} {556}},\ \bibinfo {pages}
  {74–79} (\bibinfo {year} {2018}{\natexlab{b}})}\BibitemShut {NoStop}%
\bibitem [{\citenamefont {G\'omez-Ruiz}\ \emph
  {et~al.}(2018{\natexlab{a}})\citenamefont {G\'omez-Ruiz}, \citenamefont
  {Mendoza-Arenas}, \citenamefont {Rodr\'{\i}guez}, \citenamefont {Tejedor},\
  and\ \citenamefont {Quiroga}}]{FernandoTwoTimeCorrelations}%
  \BibitemOpen
  \bibfield  {author} {\bibinfo {author} {\bibfnamefont {F.~J.}\ \bibnamefont
  {G\'omez-Ruiz}}, \bibinfo {author} {\bibfnamefont {J.~J.}\ \bibnamefont
  {Mendoza-Arenas}}, \bibinfo {author} {\bibfnamefont {F.~J.}\ \bibnamefont
  {Rodr\'{\i}guez}}, \bibinfo {author} {\bibfnamefont {C.}~\bibnamefont
  {Tejedor}},\ and\ \bibinfo {author} {\bibfnamefont {L.}~\bibnamefont
  {Quiroga}},\ }\bibfield  {title} {\bibinfo {title} {Universal two-time
  correlations, out-of-time-ordered correlators, and {L}eggett-{G}arg
  inequality violation by edge {M}ajorana fermion qubits},\ }\href
  {https://doi.org/10.1103/PhysRevB.97.235134} {\bibfield  {journal} {\bibinfo
  {journal} {Phys. Rev. B}\ }\textbf {\bibinfo {volume} {97}},\ \bibinfo
  {pages} {235134} (\bibinfo {year} {2018}{\natexlab{a}})}\BibitemShut
  {NoStop}%
\bibitem [{\citenamefont {Bermudez}\ \emph {et~al.}(2010)\citenamefont
  {Bermudez}, \citenamefont {Amico},\ and\ \citenamefont
  {Martin-Delgado}}]{DynamicalDelocalizationMajorana}%
  \BibitemOpen
  \bibfield  {author} {\bibinfo {author} {\bibfnamefont {A.}~\bibnamefont
  {Bermudez}}, \bibinfo {author} {\bibfnamefont {L.}~\bibnamefont {Amico}},\
  and\ \bibinfo {author} {\bibfnamefont {M.~A.}\ \bibnamefont
  {Martin-Delgado}},\ }\bibfield  {title} {\bibinfo {title} {Dynamical
  delocalization of {M}ajorana edge states by sweeping across a quantum
  critical point},\ }\href {http://stacks.iop.org/1367-2630/12/i=5/a=055014}
  {\bibfield  {journal} {\bibinfo  {journal} {New. J. Phys.}\ }\textbf
  {\bibinfo {volume} {12}},\ \bibinfo {pages} {055014} (\bibinfo {year}
  {2010})}\BibitemShut {NoStop}%
\bibitem [{\citenamefont {Amico}\ \emph {et~al.}(2008)\citenamefont {Amico},
  \citenamefont {Fazio}, \citenamefont {Osterloh},\ and\ \citenamefont
  {Vedral}}]{EntanglementInManyBody}%
  \BibitemOpen
  \bibfield  {author} {\bibinfo {author} {\bibfnamefont {L.}~\bibnamefont
  {Amico}}, \bibinfo {author} {\bibfnamefont {R.}~\bibnamefont {Fazio}},
  \bibinfo {author} {\bibfnamefont {A.}~\bibnamefont {Osterloh}},\ and\
  \bibinfo {author} {\bibfnamefont {V.}~\bibnamefont {Vedral}},\ }\bibfield
  {title} {\bibinfo {title} {Entanglement in many-body systems},\ }\href
  {https://doi.org/10.1103/RevModPhys.80.517} {\bibfield  {journal} {\bibinfo
  {journal} {Rev. Mod. Phys.}\ }\textbf {\bibinfo {volume} {80}},\ \bibinfo
  {pages} {517} (\bibinfo {year} {2008})}\BibitemShut {NoStop}%
\bibitem [{\citenamefont {Aasen}\ \emph {et~al.}(2016)\citenamefont {Aasen},
  \citenamefont {Hell}, \citenamefont {Mishmash}, \citenamefont {Higginbotham},
  \citenamefont {Danon}, \citenamefont {Leijnse}, \citenamefont {Jespersen},
  \citenamefont {Folk}, \citenamefont {Marcus}, \citenamefont {Flensberg},\
  and\ \citenamefont {Alicea}}]{AasenMajoranaQuantumComputing}%
  \BibitemOpen
  \bibfield  {author} {\bibinfo {author} {\bibfnamefont {D.}~\bibnamefont
  {Aasen}}, \bibinfo {author} {\bibfnamefont {M.}~\bibnamefont {Hell}},
  \bibinfo {author} {\bibfnamefont {R.~V.}\ \bibnamefont {Mishmash}}, \bibinfo
  {author} {\bibfnamefont {A.}~\bibnamefont {Higginbotham}}, \bibinfo {author}
  {\bibfnamefont {J.}~\bibnamefont {Danon}}, \bibinfo {author} {\bibfnamefont
  {M.}~\bibnamefont {Leijnse}}, \bibinfo {author} {\bibfnamefont {T.~S.}\
  \bibnamefont {Jespersen}}, \bibinfo {author} {\bibfnamefont {J.~A.}\
  \bibnamefont {Folk}}, \bibinfo {author} {\bibfnamefont {C.~M.}\ \bibnamefont
  {Marcus}}, \bibinfo {author} {\bibfnamefont {K.}~\bibnamefont {Flensberg}},\
  and\ \bibinfo {author} {\bibfnamefont {J.}~\bibnamefont {Alicea}},\
  }\bibfield  {title} {\bibinfo {title} {Milestones toward {M}ajorana-based
  quantum computing},\ }\href {https://doi.org/10.1103/PhysRevX.6.031016}
  {\bibfield  {journal} {\bibinfo  {journal} {Phys. Rev. X}\ }\textbf {\bibinfo
  {volume} {6}},\ \bibinfo {pages} {031016} (\bibinfo {year}
  {2016})}\BibitemShut {NoStop}%
\bibitem [{\citenamefont {Kitaev}(2001)}]{Kitaev}%
  \BibitemOpen
  \bibfield  {author} {\bibinfo {author} {\bibfnamefont {A.~Y.}\ \bibnamefont
  {Kitaev}},\ }\bibfield  {title} {\bibinfo {title} {Unpaired {M}ajorana
  fermions in quantum wires},\ }\href
  {http://stacks.iop.org/1063-7869/44/i=10S/a=S29} {\bibfield  {journal}
  {\bibinfo  {journal} {Sov. Phys. Usp.}\ }\textbf {\bibinfo {volume} {44}},\
  \bibinfo {pages} {131} (\bibinfo {year} {2001})}\BibitemShut {NoStop}%
\bibitem [{\citenamefont {Wilczek}(2009)}]{majoranareturns}%
  \BibitemOpen
  \bibfield  {author} {\bibinfo {author} {\bibfnamefont {F.}~\bibnamefont
  {Wilczek}},\ }\bibfield  {title} {\bibinfo {title} {{M}ajorana returns},\
  }\href {https://doi.org/10.1038/nphys1380} {\bibfield  {journal} {\bibinfo
  {journal} {Nat. Phys.}\ }\textbf {\bibinfo {volume} {5}},\ \bibinfo {pages}
  {614} (\bibinfo {year} {2009})}\BibitemShut {NoStop}%
\bibitem [{\citenamefont {Elliott}\ and\ \citenamefont
  {Franz}(2015)}]{majorana}%
  \BibitemOpen
  \bibfield  {author} {\bibinfo {author} {\bibfnamefont {S.~R.}\ \bibnamefont
  {Elliott}}\ and\ \bibinfo {author} {\bibfnamefont {M.}~\bibnamefont
  {Franz}},\ }\bibfield  {title} {\bibinfo {title} {Colloquium},\ }\href
  {https://doi.org/10.1103/RevModPhys.87.137} {\bibfield  {journal} {\bibinfo
  {journal} {Rev. Mod. Phys.}\ }\textbf {\bibinfo {volume} {87}},\ \bibinfo
  {pages} {137} (\bibinfo {year} {2015})}\BibitemShut {NoStop}%
\bibitem [{\citenamefont {Burton}()}]{AnyonsBurton}%
  \BibitemOpen
  \bibfield  {author} {\bibinfo {author} {\bibfnamefont {S.}~\bibnamefont
  {Burton}},\ }\href@noop {} {}\Eprint {https://arxiv.org/abs/1610.05384}
  {arXiv:1610.05384} \BibitemShut {NoStop}%
\bibitem [{\citenamefont {Wang}(2018)}]{SurvivalMajorana}%
  \BibitemOpen
  \bibfield  {author} {\bibinfo {author} {\bibfnamefont {Y.}~\bibnamefont
  {Wang}},\ }\bibfield  {title} {\bibinfo {title} {Detecting topological phases
  via survival probabilities of edge {M}ajorana fermions},\ }\href
  {https://link.aps.org/doi/10.1103/PhysRevE.98.042128} {\bibfield  {journal}
  {\bibinfo  {journal} {Phys. Rev. E}\ }\textbf {\bibinfo {volume} {98}},\
  \bibinfo {pages} {042128} (\bibinfo {year} {2018})}\BibitemShut {NoStop}%
\bibitem [{\citenamefont {Yang}\ \emph {et~al.}(2019)\citenamefont {Yang},
  \citenamefont {Iadecola}, \citenamefont {Chamon},\ and\ \citenamefont
  {Mudry}}]{ProgramableMajoranas}%
  \BibitemOpen
  \bibfield  {author} {\bibinfo {author} {\bibfnamefont {Z.-C.}\ \bibnamefont
  {Yang}}, \bibinfo {author} {\bibfnamefont {T.}~\bibnamefont {Iadecola}},
  \bibinfo {author} {\bibfnamefont {C.}~\bibnamefont {Chamon}},\ and\ \bibinfo
  {author} {\bibfnamefont {C.}~\bibnamefont {Mudry}},\ }\bibfield  {title}
  {\bibinfo {title} {Hierarchical {M}ajoranas in a programmable nanowire
  network},\ }\href {https://doi.org/10.1103/PhysRevB.99.155138} {\bibfield
  {journal} {\bibinfo  {journal} {Phys. Rev. B}\ }\textbf {\bibinfo {volume}
  {99}},\ \bibinfo {pages} {155138} (\bibinfo {year} {2019})}\BibitemShut
  {NoStop}%
\bibitem [{\citenamefont {Kim}\ \emph {et~al.}(2018)\citenamefont {Kim},
  \citenamefont {Shin}, \citenamefont {Kim}, \citenamefont {Song},\ and\
  \citenamefont {Doh}}]{andreevKim}%
  \BibitemOpen
  \bibfield  {author} {\bibinfo {author} {\bibfnamefont {N.-H.}\ \bibnamefont
  {Kim}}, \bibinfo {author} {\bibfnamefont {Y.-S.}\ \bibnamefont {Shin}},
  \bibinfo {author} {\bibfnamefont {H.-S.}\ \bibnamefont {Kim}}, \bibinfo
  {author} {\bibfnamefont {J.-D.}\ \bibnamefont {Song}},\ and\ \bibinfo
  {author} {\bibfnamefont {Y.-J.}\ \bibnamefont {Doh}},\ }\bibfield  {title}
  {\bibinfo {title} {Zero bias conductance peak in inas nanowire coupled to
  superconducting electrodes},\ }\href
  {https://doi.org/https://doi.org/10.1016/j.cap.2018.01.016} {\bibfield
  {journal} {\bibinfo  {journal} {C. Appl. Phys.}\ }\textbf {\bibinfo {volume}
  {18}},\ \bibinfo {pages} {384 } (\bibinfo {year} {2018})}\BibitemShut
  {NoStop}%
\bibitem [{\citenamefont {Trif}\ and\ \citenamefont
  {Tserkovnyak}(2012)}]{MirceaTrifHamiltonian}%
  \BibitemOpen
  \bibfield  {author} {\bibinfo {author} {\bibfnamefont {M.}~\bibnamefont
  {Trif}}\ and\ \bibinfo {author} {\bibfnamefont {Y.}~\bibnamefont
  {Tserkovnyak}},\ }\bibfield  {title} {\bibinfo {title} {Resonantly tunable
  {M}ajorana polariton in a microwave cavity},\ }\href
  {https://doi.org/10.1103/PhysRevLett.109.257002} {\bibfield  {journal}
  {\bibinfo  {journal} {Phys. Rev. Lett.}\ }\textbf {\bibinfo {volume} {109}},\
  \bibinfo {pages} {257002} (\bibinfo {year} {2012})}\BibitemShut {NoStop}%
\bibitem [{\citenamefont {Trif}\ and\ \citenamefont
  {Simon}(2019)}]{trif2019prl}%
  \BibitemOpen
  \bibfield  {author} {\bibinfo {author} {\bibfnamefont {M.}~\bibnamefont
  {Trif}}\ and\ \bibinfo {author} {\bibfnamefont {P.}~\bibnamefont {Simon}},\
  }\bibfield  {title} {\bibinfo {title} {Braiding of {M}ajorana fermions in a
  cavity},\ }\href {https://doi.org/10.1103/PhysRevLett.122.236803} {\bibfield
  {journal} {\bibinfo  {journal} {Phys. Rev. Lett.}\ }\textbf {\bibinfo
  {volume} {122}},\ \bibinfo {pages} {236803} (\bibinfo {year}
  {2019})}\BibitemShut {NoStop}%
\bibitem [{\citenamefont {Dmytruk}\ \emph {et~al.}(2015)\citenamefont
  {Dmytruk}, \citenamefont {Trif},\ and\ \citenamefont
  {Simon}}]{TrifMajoranasSpinOrbit}%
  \BibitemOpen
  \bibfield  {author} {\bibinfo {author} {\bibfnamefont {O.}~\bibnamefont
  {Dmytruk}}, \bibinfo {author} {\bibfnamefont {M.}~\bibnamefont {Trif}},\ and\
  \bibinfo {author} {\bibfnamefont {P.}~\bibnamefont {Simon}},\ }\bibfield
  {title} {\bibinfo {title} {Cavity quantum electrodynamics with mesoscopic
  topological superconductors},\ }\href
  {https://doi.org/10.1103/PhysRevB.92.245432} {\bibfield  {journal} {\bibinfo
  {journal} {Phys. Rev. B}\ }\textbf {\bibinfo {volume} {92}},\ \bibinfo
  {pages} {245432} (\bibinfo {year} {2015})}\BibitemShut {NoStop}%
\bibitem [{\citenamefont {Acevedo}\ \emph
  {et~al.}(2015{\natexlab{a}})\citenamefont {Acevedo}, \citenamefont {Quiroga},
  \citenamefont {Rodr\'{\i}guez},\ and\ \citenamefont
  {Johnson}}]{Acevedo2_2015}%
  \BibitemOpen
  \bibfield  {author} {\bibinfo {author} {\bibfnamefont {O.~L.}\ \bibnamefont
  {Acevedo}}, \bibinfo {author} {\bibfnamefont {L.}~\bibnamefont {Quiroga}},
  \bibinfo {author} {\bibfnamefont {F.~J.}\ \bibnamefont {Rodr\'{\i}guez}},\
  and\ \bibinfo {author} {\bibfnamefont {N.~F.}\ \bibnamefont {Johnson}},\
  }\bibfield  {title} {\bibinfo {title} {Large dynamic light-matter
  entanglement from driving neither too fast nor too slow},\ }\href
  {https://doi.org/10.1103/PhysRevA.92.032330} {\bibfield  {journal} {\bibinfo
  {journal} {Phys. Rev. A}\ }\textbf {\bibinfo {volume} {92}},\ \bibinfo
  {pages} {032330} (\bibinfo {year} {2015}{\natexlab{a}})}\BibitemShut
  {NoStop}%
\bibitem [{\citenamefont {Acevedo}\ \emph
  {et~al.}(2015{\natexlab{b}})\citenamefont {Acevedo}, \citenamefont {Quiroga},
  \citenamefont {Rodr{\'{\i}}guez},\ and\ \citenamefont
  {Johnson}}]{Acevedo_2015}%
  \BibitemOpen
  \bibfield  {author} {\bibinfo {author} {\bibfnamefont {O.~L.}\ \bibnamefont
  {Acevedo}}, \bibinfo {author} {\bibfnamefont {L.}~\bibnamefont {Quiroga}},
  \bibinfo {author} {\bibfnamefont {F.~J.}\ \bibnamefont {Rodr{\'{\i}}guez}},\
  and\ \bibinfo {author} {\bibfnamefont {N.~F.}\ \bibnamefont {Johnson}},\
  }\bibfield  {title} {\bibinfo {title} {Robust quantum correlations in
  out-of-equilibrium matter{\textendash}light systems},\ }\href
  {https://doi.org/10.1088/1367-2630/17/9/093005} {\bibfield  {journal}
  {\bibinfo  {journal} {New J. Phys.}\ }\textbf {\bibinfo {volume} {17}},\
  \bibinfo {pages} {093005} (\bibinfo {year} {2015}{\natexlab{b}})}\BibitemShut
  {NoStop}%
\bibitem [{\citenamefont {G\'omez-Ruiz}\ \emph
  {et~al.}(2018{\natexlab{b}})\citenamefont {G\'omez-Ruiz}, \citenamefont
  {Acevedo}, \citenamefont {Rodr\'iguez}, \citenamefont {Quiroga},\ and\
  \citenamefont {Johnson}}]{Gomez2018}%
  \BibitemOpen
  \bibfield  {author} {\bibinfo {author} {\bibfnamefont {F.~J.}\ \bibnamefont
  {G\'omez-Ruiz}}, \bibinfo {author} {\bibfnamefont {O.~L.}\ \bibnamefont
  {Acevedo}}, \bibinfo {author} {\bibfnamefont {F.~J.}\ \bibnamefont
  {Rodr\'iguez}}, \bibinfo {author} {\bibfnamefont {L.}~\bibnamefont
  {Quiroga}},\ and\ \bibinfo {author} {\bibfnamefont {N.~F.}\ \bibnamefont
  {Johnson}},\ }\bibfield  {title} {\bibinfo {title} {Pulsed generation of
  quantum coherences and non-classicality in light-matter systems},\ }\href
  {https://doi.org/10.3389/fphy.2018.00092} {\bibfield  {journal} {\bibinfo
  {journal} {Front. Phys.}\ }\textbf {\bibinfo {volume} {6}},\ \bibinfo {pages}
  {92} (\bibinfo {year} {2018}{\natexlab{b}})}\BibitemShut {NoStop}%
\bibitem [{\citenamefont {Sachdev}(1998)}]{qpt}%
  \BibitemOpen
  \bibfield  {author} {\bibinfo {author} {\bibfnamefont {S.}~\bibnamefont
  {Sachdev}},\ }\href@noop {} {\emph {\bibinfo {title} {Quantum phase
  transitions}}}\ (\bibinfo  {publisher} {Cambridge University Press.},\
  \bibinfo {year} {1998})\BibitemShut {NoStop}%
\bibitem [{\citenamefont {Suzuki}\ \emph {et~al.}(2012)\citenamefont {Suzuki},
  \citenamefont {Inoue},\ and\ \citenamefont {Chakrabarti}}]{susuki}%
  \BibitemOpen
  \bibfield  {author} {\bibinfo {author} {\bibfnamefont {S.}~\bibnamefont
  {Suzuki}}, \bibinfo {author} {\bibfnamefont {J.}~\bibnamefont {Inoue}},\ and\
  \bibinfo {author} {\bibfnamefont {B.}~\bibnamefont {Chakrabarti}},\ }\href
  {https://books.google.com.co/books?id=y1S5BQAAQBAJ} {\emph {\bibinfo {title}
  {Quantum {I}sing Phases and Transitions in Transverse {I}sing Models}}},\
  Lecture Notes in Physics\ (\bibinfo  {publisher} {Springer Berlin
  Heidelberg},\ \bibinfo {year} {2012})\BibitemShut {NoStop}%
\bibitem [{\citenamefont {Cortese}\ \emph {et~al.}(2017)\citenamefont
  {Cortese}, \citenamefont {Garziano},\ and\ \citenamefont
  {De~Liberato}}]{PolaritonLiberato}%
  \BibitemOpen
  \bibfield  {author} {\bibinfo {author} {\bibfnamefont {E.}~\bibnamefont
  {Cortese}}, \bibinfo {author} {\bibfnamefont {L.}~\bibnamefont {Garziano}},\
  and\ \bibinfo {author} {\bibfnamefont {S.}~\bibnamefont {De~Liberato}},\
  }\bibfield  {title} {\bibinfo {title} {Polariton spectrum of the
  {D}icke-{I}sing model},\ }\href {https://doi.org/10.1103/PhysRevA.96.053861}
  {\bibfield  {journal} {\bibinfo  {journal} {Phys. Rev. A}\ }\textbf {\bibinfo
  {volume} {96}},\ \bibinfo {pages} {053861} (\bibinfo {year}
  {2017})}\BibitemShut {NoStop}%
\bibitem [{\citenamefont {Greiter}\ \emph {et~al.}(2014)\citenamefont
  {Greiter}, \citenamefont {Schnells},\ and\ \citenamefont
  {Thomale}}]{Ising-Kitaev}%
  \BibitemOpen
  \bibfield  {author} {\bibinfo {author} {\bibfnamefont {M.}~\bibnamefont
  {Greiter}}, \bibinfo {author} {\bibfnamefont {V.}~\bibnamefont {Schnells}},\
  and\ \bibinfo {author} {\bibfnamefont {R.}~\bibnamefont {Thomale}},\
  }\bibfield  {title} {\bibinfo {title} {The {1D} {I}sing model and the
  topological phase of the {K}itaev chain},\ }\href
  {https://doi.org/https://doi.org/10.1016/j.aop.2014.08.013} {\bibfield
  {journal} {\bibinfo  {journal} {Ann. Phys.}\ }\textbf {\bibinfo {volume}
  {351}},\ \bibinfo {pages} {1026 } (\bibinfo {year} {2014})}\BibitemShut
  {NoStop}%
\bibitem [{\citenamefont {Schollwöck}(2011)}]{SCHOLLWOCK}%
  \BibitemOpen
  \bibfield  {author} {\bibinfo {author} {\bibfnamefont {U.}~\bibnamefont
  {Schollwöck}},\ }\bibfield  {title} {\bibinfo {title} {The density-matrix
  renormalization group in the age of matrix product states},\ }\href
  {https://doi.org/https://doi.org/10.1016/j.aop.2010.09.012} {\bibfield
  {journal} {\bibinfo  {journal} {Ann. Phys.}\ }\textbf {\bibinfo {volume}
  {326}},\ \bibinfo {pages} {96 } (\bibinfo {year} {2011})}\BibitemShut
  {NoStop}%
\bibitem [{\citenamefont {Or\'us}(2019)}]{Or_s_2019DMRG}%
  \BibitemOpen
  \bibfield  {author} {\bibinfo {author} {\bibfnamefont {R.}~\bibnamefont
  {Or\'us}},\ }\bibfield  {title} {\bibinfo {title} {Tensor networks for
  complex quantum systems},\ }\href {https://doi.org/10.1038/s42254-019-0086-7}
  {\bibfield  {journal} {\bibinfo  {journal} {Nat. Rev. Phys.}\ }\textbf
  {\bibinfo {volume} {1}},\ \bibinfo {pages} {538–550} (\bibinfo {year}
  {2019})}\BibitemShut {NoStop}%
\bibitem [{\citenamefont {S.~Al-Assam}\ and\ \citenamefont {team}(2016)}]{TNT}%
  \BibitemOpen
  \bibfield  {author} {\bibinfo {author} {\bibfnamefont {D.~J.}\ \bibnamefont
  {S.~Al-Assam}, \bibfnamefont {S.~R.~Clark}}\ and\ \bibinfo {author}
  {\bibfnamefont {T.~D.}\ \bibnamefont {team}},\ }\href@noop {} {\bibinfo
  {title} {Tensor network theory library, beta version 1.2.0}},\ \bibinfo
  {howpublished} {\url{http://www.tensornetworktheory.org/}} (\bibinfo {year}
  {2016})\BibitemShut {NoStop}%
\bibitem [{\citenamefont {Al-Assam}\ \emph {et~al.}(2017)\citenamefont
  {Al-Assam}, \citenamefont {Clark},\ and\ \citenamefont {Jaksch}}]{Sarah}%
  \BibitemOpen
  \bibfield  {author} {\bibinfo {author} {\bibfnamefont {S.}~\bibnamefont
  {Al-Assam}}, \bibinfo {author} {\bibfnamefont {S.~R.}\ \bibnamefont
  {Clark}},\ and\ \bibinfo {author} {\bibfnamefont {D.}~\bibnamefont
  {Jaksch}},\ }\bibfield  {title} {\bibinfo {title} {The tensor network theory
  library},\ }\href {http://stacks.iop.org/1742-5468/2017/i=9/a=093102}
  {\bibfield  {journal} {\bibinfo  {journal} {J. Stat. Mech.}\ }\textbf
  {\bibinfo {volume} {2017}},\ \bibinfo {pages} {093102} (\bibinfo {year}
  {2017})}\BibitemShut {NoStop}%
\bibitem [{\citenamefont {Gammelmark}\ and\ \citenamefont
  {M\o{}lmer}(2012)}]{GandMDMRG}%
  \BibitemOpen
  \bibfield  {author} {\bibinfo {author} {\bibfnamefont {S.}~\bibnamefont
  {Gammelmark}}\ and\ \bibinfo {author} {\bibfnamefont {K.}~\bibnamefont
  {M\o{}lmer}},\ }\bibfield  {title} {\bibinfo {title} {Interacting spins in a
  cavity: Finite-size effects and symmetry-breaking dynamics},\ }\href
  {https://doi.org/10.1103/PhysRevA.85.042114} {\bibfield  {journal} {\bibinfo
  {journal} {Phys. Rev. A}\ }\textbf {\bibinfo {volume} {85}},\ \bibinfo
  {pages} {042114} (\bibinfo {year} {2012})}\BibitemShut {NoStop}%
\bibitem [{\citenamefont {Halati}\ \emph {et~al.}()\citenamefont {Halati},
  \citenamefont {Sheikhan},\ and\ \citenamefont {Kollath}}]{CavityKollath}%
  \BibitemOpen
  \bibfield  {author} {\bibinfo {author} {\bibfnamefont {C.-M.}\ \bibnamefont
  {Halati}}, \bibinfo {author} {\bibfnamefont {A.}~\bibnamefont {Sheikhan}},\
  and\ \bibinfo {author} {\bibfnamefont {C.}~\bibnamefont {Kollath}},\
  }\href@noop {} {}\Eprint {https://arxiv.org/abs/2004.11807}
  {arXiv:2004.11807} \BibitemShut {NoStop}%
\bibitem [{\citenamefont {Wolf}\ \emph {et~al.}(2014)\citenamefont {Wolf},
  \citenamefont {McCulloch},\ and\ \citenamefont {Schollw\"ock}}]{wolf2014prb}%
  \BibitemOpen
  \bibfield  {author} {\bibinfo {author} {\bibfnamefont {F.~A.}\ \bibnamefont
  {Wolf}}, \bibinfo {author} {\bibfnamefont {I.~P.}\ \bibnamefont
  {McCulloch}},\ and\ \bibinfo {author} {\bibfnamefont {U.}~\bibnamefont
  {Schollw\"ock}},\ }\bibfield  {title} {\bibinfo {title} {Solving
  nonequilibrium dynamical mean-field theory using matrix product states},\
  }\href {https://doi.org/10.1103/PhysRevB.90.235131} {\bibfield  {journal}
  {\bibinfo  {journal} {Phys. Rev. B}\ }\textbf {\bibinfo {volume} {90}},\
  \bibinfo {pages} {235131} (\bibinfo {year} {2014})}\BibitemShut {NoStop}%
\bibitem [{\citenamefont {Mendoza-Arenas}\ \emph {et~al.}(2017)\citenamefont
  {Mendoza-Arenas}, \citenamefont {G\'omez-Ruiz}, \citenamefont {Eckstein},
  \citenamefont {Jaksch},\ and\ \citenamefont {Clark}}]{Mendoza:2017}%
  \BibitemOpen
  \bibfield  {author} {\bibinfo {author} {\bibfnamefont {J.~J.}\ \bibnamefont
  {Mendoza-Arenas}}, \bibinfo {author} {\bibfnamefont {F.~J.}\ \bibnamefont
  {G\'omez-Ruiz}}, \bibinfo {author} {\bibfnamefont {M.}~\bibnamefont
  {Eckstein}}, \bibinfo {author} {\bibfnamefont {D.}~\bibnamefont {Jaksch}},\
  and\ \bibinfo {author} {\bibfnamefont {S.~R.}\ \bibnamefont {Clark}},\
  }\bibfield  {title} {\bibinfo {title} {Ultra-fast control of magnetic
  relaxation in a periodically driven {H}ubbard model},\ }\href
  {https://doi.org/10.1002/andp.201700024} {\bibfield  {journal} {\bibinfo
  {journal} {Ann. Phys. (Berlin)}\ }\textbf {\bibinfo {volume} {529}},\
  \bibinfo {pages} {1700024} (\bibinfo {year} {2017})}\BibitemShut {NoStop}%
\bibitem [{\citenamefont {Rams}\ and\ \citenamefont
  {Zwolak}(2020)}]{zwolak2020prl}%
  \BibitemOpen
  \bibfield  {author} {\bibinfo {author} {\bibfnamefont {M.~M.}\ \bibnamefont
  {Rams}}\ and\ \bibinfo {author} {\bibfnamefont {M.}~\bibnamefont {Zwolak}},\
  }\bibfield  {title} {\bibinfo {title} {Breaking the entanglement barrier:
  Tensor network simulation of quantum transport},\ }\href
  {https://doi.org/10.1103/PhysRevLett.124.137701} {\bibfield  {journal}
  {\bibinfo  {journal} {Phys. Rev. Lett.}\ }\textbf {\bibinfo {volume} {124}},\
  \bibinfo {pages} {137701} (\bibinfo {year} {2020})}\BibitemShut {NoStop}%
\bibitem [{\citenamefont {Brenes}\ \emph {et~al.}(2020)\citenamefont {Brenes},
  \citenamefont {Mendoza-Arenas}, \citenamefont {Purkayastha}, \citenamefont
  {Mitchison}, \citenamefont {Clark},\ and\ \citenamefont
  {Goold}}]{brenes2020}%
  \BibitemOpen
  \bibfield  {author} {\bibinfo {author} {\bibfnamefont {M.}~\bibnamefont
  {Brenes}}, \bibinfo {author} {\bibfnamefont {J.~J.}\ \bibnamefont
  {Mendoza-Arenas}}, \bibinfo {author} {\bibfnamefont {A.}~\bibnamefont
  {Purkayastha}}, \bibinfo {author} {\bibfnamefont {M.~T.}\ \bibnamefont
  {Mitchison}}, \bibinfo {author} {\bibfnamefont {S.~R.}\ \bibnamefont
  {Clark}},\ and\ \bibinfo {author} {\bibfnamefont {J.}~\bibnamefont {Goold}},\
  }\bibfield  {title} {\bibinfo {title} {Tensor-network method to simulate
  strongly interacting quantum thermal machines},\ }\href
  {https://doi.org/10.1103/PhysRevX.10.031040} {\bibfield  {journal} {\bibinfo
  {journal} {Phys. Rev. X}\ }\textbf {\bibinfo {volume} {10}},\ \bibinfo
  {pages} {031040} (\bibinfo {year} {2020})}\BibitemShut {NoStop}%
\bibitem [{\citenamefont {Reslen}(2018)}]{ReslenQ}%
  \BibitemOpen
  \bibfield  {author} {\bibinfo {author} {\bibfnamefont {J.}~\bibnamefont
  {Reslen}},\ }\bibfield  {title} {\bibinfo {title} {End-to-end correlations in
  the {K}itaev chain},\ }\href {https://doi.org/10.1088/2399-6528/aae4c5}
  {\bibfield  {journal} {\bibinfo  {journal} {J. Phys. Comm.}\ }\textbf
  {\bibinfo {volume} {2}},\ \bibinfo {pages} {105006} (\bibinfo {year}
  {2018})}\BibitemShut {NoStop}%
\bibitem [{\citenamefont {Lee}\ \emph {et~al.}(2016)\citenamefont {Lee},
  \citenamefont {Han},\ and\ \citenamefont {Choi}}]{LeeQ}%
  \BibitemOpen
  \bibfield  {author} {\bibinfo {author} {\bibfnamefont {M.}~\bibnamefont
  {Lee}}, \bibinfo {author} {\bibfnamefont {S.}~\bibnamefont {Han}},\ and\
  \bibinfo {author} {\bibfnamefont {M.-S.}\ \bibnamefont {Choi}},\ }\bibfield
  {title} {\bibinfo {title} {Spatiotemporal evolution of topological order upon
  quantum quench across the critical point},\ }\href
  {https://doi.org/10.1088/1367-2630/18/6/063004} {\bibfield  {journal}
  {\bibinfo  {journal} {New J. Phys.}\ }\textbf {\bibinfo {volume} {18}},\
  \bibinfo {pages} {063004} (\bibinfo {year} {2016})}\BibitemShut {NoStop}%
\bibitem [{\citenamefont {Vidal}(2004)}]{Vidal}%
  \BibitemOpen
  \bibfield  {author} {\bibinfo {author} {\bibfnamefont {G.}~\bibnamefont
  {Vidal}},\ }\bibfield  {title} {\bibinfo {title} {Efficient simulation of
  one-dimensional quantum many-body systems},\ }\href
  {https://doi.org/10.1103/PhysRevLett.93.040502} {\bibfield  {journal}
  {\bibinfo  {journal} {Phys. Rev. Lett.}\ }\textbf {\bibinfo {volume} {93}},\
  \bibinfo {pages} {040502} (\bibinfo {year} {2004})}\BibitemShut {NoStop}%
\bibitem [{\citenamefont {Olivares}\ \emph {et~al.}(2018)\citenamefont
  {Olivares}, \citenamefont {Cialdi},\ and\ \citenamefont
  {Paris}}]{GaussianExpectedOlivares}%
  \BibitemOpen
  \bibfield  {author} {\bibinfo {author} {\bibfnamefont {S.}~\bibnamefont
  {Olivares}}, \bibinfo {author} {\bibfnamefont {S.}~\bibnamefont {Cialdi}},\
  and\ \bibinfo {author} {\bibfnamefont {M.~G.}\ \bibnamefont {Paris}},\
  }\bibfield  {title} {\bibinfo {title} {Homodyning the g(2)(0) of {G}aussian
  states},\ }\href {https://doi.org/10.1016/j.optcom.2018.05.090} {\bibfield
  {journal} {\bibinfo  {journal} {Opt. Commun.}\ }\textbf {\bibinfo {volume}
  {426}},\ \bibinfo {pages} {547–552} (\bibinfo {year} {2018})}\BibitemShut
  {NoStop}%
\bibitem [{\citenamefont {Park}(2018)}]{GaussianRenyiDaeKil}%
  \BibitemOpen
  \bibfield  {author} {\bibinfo {author} {\bibfnamefont {D.}~\bibnamefont
  {Park}},\ }\bibfield  {title} {\bibinfo {title} {Dynamics of entanglement and
  uncertainty relation in coupled harmonic oscillator system: exact result},\
  }\href {https://doi.org/10.1007/s11128-018-1914-x} {\bibfield  {journal}
  {\bibinfo  {journal} {Quant. Infor. Proc.}\ }\textbf {\bibinfo {volume}
  {17}},\ \bibinfo {pages} {147} (\bibinfo {year} {2018})}\BibitemShut
  {NoStop}%
\bibitem [{\citenamefont {Genoni}\ and\ \citenamefont
  {Paris}(2010)}]{GaussianVonNeumann}%
  \BibitemOpen
  \bibfield  {author} {\bibinfo {author} {\bibfnamefont {M.~G.}\ \bibnamefont
  {Genoni}}\ and\ \bibinfo {author} {\bibfnamefont {M.~G.~A.}\ \bibnamefont
  {Paris}},\ }\bibfield  {title} {\bibinfo {title} {Quantifying
  non-{G}aussianity for quantum information},\ }\href
  {https://doi.org/10.1103/PhysRevA.82.052341} {\bibfield  {journal} {\bibinfo
  {journal} {Phys. Rev. A}\ }\textbf {\bibinfo {volume} {82}},\ \bibinfo
  {pages} {052341} (\bibinfo {year} {2010})}\BibitemShut {NoStop}%
\bibitem [{\citenamefont {Alexanian}(2015)}]{GaussianAlexanian}%
  \BibitemOpen
  \bibfield  {author} {\bibinfo {author} {\bibfnamefont {M.}~\bibnamefont
  {Alexanian}},\ }\bibfield  {title} {\bibinfo {title} {Temporal second-order
  coherence function for displaced-squeezed thermal states},\ }\href
  {https://doi.org/10.1080/09500340.2015.1112441} {\bibfield  {journal}
  {\bibinfo  {journal} {J. Mod. Opt.}\ }\textbf {\bibinfo {volume} {63}},\
  \bibinfo {pages} {961–967} (\bibinfo {year} {2015})}\BibitemShut {NoStop}%
\bibitem [{\citenamefont {Alexanian}(2018)}]{GaussianAlexanianExpected}%
  \BibitemOpen
  \bibfield  {author} {\bibinfo {author} {\bibfnamefont {M.}~\bibnamefont
  {Alexanian}},\ }\bibfield  {title} {\bibinfo {title} {Non-classicality
  criteria: {G}lauber-{S}udarshan {P} function and mandel {$Q_M$} parameter},\
  }\href {https://doi.org/10.1080/09500340.2017.1374481} {\bibfield  {journal}
  {\bibinfo  {journal} {J. Mod. Opt.}\ }\textbf {\bibinfo {volume} {65}},\
  \bibinfo {pages} {16} (\bibinfo {year} {2018})}\BibitemShut {NoStop}%
\bibitem [{\citenamefont {Adesso}\ \emph {et~al.}(2012)\citenamefont {Adesso},
  \citenamefont {Girolami},\ and\ \citenamefont {Serafini}}]{adessoPRL2012}%
  \BibitemOpen
  \bibfield  {author} {\bibinfo {author} {\bibfnamefont {G.}~\bibnamefont
  {Adesso}}, \bibinfo {author} {\bibfnamefont {D.}~\bibnamefont {Girolami}},\
  and\ \bibinfo {author} {\bibfnamefont {A.}~\bibnamefont {Serafini}},\
  }\bibfield  {title} {\bibinfo {title} {{Measuring {G}aussian Quantum
  Information and Correlations Using the R\'enyi Entropy of Order 2}},\ }\href
  {https://doi.org/10.1103/PhysRevLett.109.190502} {\bibfield  {journal}
  {\bibinfo  {journal} {Phys. Rev. Lett.}\ }\textbf {\bibinfo {volume} {109}},\
  \bibinfo {pages} {190502} (\bibinfo {year} {2012})}\BibitemShut {NoStop}%
\bibitem [{\citenamefont {Haus}(2004)}]{Haus_2004squeezed}%
  \BibitemOpen
  \bibfield  {author} {\bibinfo {author} {\bibfnamefont {H.~A.}\ \bibnamefont
  {Haus}},\ }\bibfield  {title} {\bibinfo {title} {Quantum noise, quantum
  measurement, and squeezing},\ }\href
  {https://doi.org/10.1088/1464-4266/6/8/001} {\bibfield  {journal} {\bibinfo
  {journal} {J. Opt. B}\ }\textbf {\bibinfo {volume} {6}},\ \bibinfo {pages}
  {S626} (\bibinfo {year} {2004})}\BibitemShut {NoStop}%
\bibitem [{\citenamefont {Andrews}(2015)}]{andrews2015photonics}%
  \BibitemOpen
  \bibfield  {author} {\bibinfo {author} {\bibfnamefont {D.}~\bibnamefont
  {Andrews}},\ }\href {https://books.google.com.co/books?id=muY1BgAAQBAJ}
  {\emph {\bibinfo {title} {Photonics, Volume 1: Fundamentals of Photonics and
  Physics}}},\ A Wiley-Science Wise Co-Publication\ (\bibinfo  {publisher}
  {Wiley},\ \bibinfo {year} {2015})\BibitemShut {NoStop}%
\bibitem [{\citenamefont {Eichler}\ \emph {et~al.}(2011)\citenamefont
  {Eichler}, \citenamefont {Bozyigit}, \citenamefont {Lang}, \citenamefont
  {Baur}, \citenamefont {Steffen}, \citenamefont {Fink}, \citenamefont
  {Filipp},\ and\ \citenamefont {Wallraff}}]{2modeSqueezingWallraff}%
  \BibitemOpen
  \bibfield  {author} {\bibinfo {author} {\bibfnamefont {C.}~\bibnamefont
  {Eichler}}, \bibinfo {author} {\bibfnamefont {D.}~\bibnamefont {Bozyigit}},
  \bibinfo {author} {\bibfnamefont {C.}~\bibnamefont {Lang}}, \bibinfo {author}
  {\bibfnamefont {M.}~\bibnamefont {Baur}}, \bibinfo {author} {\bibfnamefont
  {L.}~\bibnamefont {Steffen}}, \bibinfo {author} {\bibfnamefont {J.~M.}\
  \bibnamefont {Fink}}, \bibinfo {author} {\bibfnamefont {S.}~\bibnamefont
  {Filipp}},\ and\ \bibinfo {author} {\bibfnamefont {A.}~\bibnamefont
  {Wallraff}},\ }\bibfield  {title} {\bibinfo {title} {Observation of two-mode
  squeezing in the microwave frequency domain},\ }\href
  {https://doi.org/10.1103/PhysRevLett.107.113601} {\bibfield  {journal}
  {\bibinfo  {journal} {Phys. Rev. Lett.}\ }\textbf {\bibinfo {volume} {107}},\
  \bibinfo {pages} {113601} (\bibinfo {year} {2011})}\BibitemShut {NoStop}%
\bibitem [{\citenamefont {Lang}\ \emph {et~al.}(2013)\citenamefont {Lang},
  \citenamefont {Eichler}, \citenamefont {Steffen}, \citenamefont {Fink},
  \citenamefont {Woolley}, \citenamefont {Blais},\ and\ \citenamefont
  {Wallraff}}]{LangG2}%
  \BibitemOpen
  \bibfield  {author} {\bibinfo {author} {\bibfnamefont {C.}~\bibnamefont
  {Lang}}, \bibinfo {author} {\bibfnamefont {C.}~\bibnamefont {Eichler}},
  \bibinfo {author} {\bibfnamefont {L.}~\bibnamefont {Steffen}}, \bibinfo
  {author} {\bibfnamefont {J.~M.}\ \bibnamefont {Fink}}, \bibinfo {author}
  {\bibfnamefont {M.~J.}\ \bibnamefont {Woolley}}, \bibinfo {author}
  {\bibfnamefont {A.}~\bibnamefont {Blais}},\ and\ \bibinfo {author}
  {\bibfnamefont {A.}~\bibnamefont {Wallraff}},\ }\bibfield  {title} {\bibinfo
  {title} {Correlations, indistinguishability and entanglement in
  {H}ong-{O}u-{M}andel experiments at microwave frequencies},\ }\href
  {https://doi.org/10.1038/nphys2612} {\bibfield  {journal} {\bibinfo
  {journal} {Nature Physics}\ }\textbf {\bibinfo {volume} {9}},\ \bibinfo
  {pages} {345–348} (\bibinfo {year} {2013})}\BibitemShut {NoStop}%
\bibitem [{\citenamefont {Vidal}\ \emph {et~al.}(2003)\citenamefont {Vidal},
  \citenamefont {Latorre}, \citenamefont {Rico},\ and\ \citenamefont
  {Kitaev}}]{EntanglementAtCriticality}%
  \BibitemOpen
  \bibfield  {author} {\bibinfo {author} {\bibfnamefont {G.}~\bibnamefont
  {Vidal}}, \bibinfo {author} {\bibfnamefont {J.~I.}\ \bibnamefont {Latorre}},
  \bibinfo {author} {\bibfnamefont {E.}~\bibnamefont {Rico}},\ and\ \bibinfo
  {author} {\bibfnamefont {A.}~\bibnamefont {Kitaev}},\ }\bibfield  {title}
  {\bibinfo {title} {Entanglement in quantum critical phenomena},\ }\href
  {https://doi.org/10.1103/PhysRevLett.90.227902} {\bibfield  {journal}
  {\bibinfo  {journal} {Phys. Rev. Lett.}\ }\textbf {\bibinfo {volume} {90}},\
  \bibinfo {pages} {227902} (\bibinfo {year} {2003})}\BibitemShut {NoStop}%
\bibitem [{\citenamefont {Gammelmark}\ and\ \citenamefont
  {Mølmer}(2011)}]{G&MMeanField}%
  \BibitemOpen
  \bibfield  {author} {\bibinfo {author} {\bibfnamefont {S.}~\bibnamefont
  {Gammelmark}}\ and\ \bibinfo {author} {\bibfnamefont {K.}~\bibnamefont
  {Mølmer}},\ }\bibfield  {title} {\bibinfo {title} {Phase transitions and
  {H}eisenberg limited metrology in an {I}sing chain interacting with a
  single-mode cavity field},\ }\href
  {http://stacks.iop.org/1367-2630/13/i=5/a=053035} {\bibfield  {journal}
  {\bibinfo  {journal} {New J. Phys.}\ }\textbf {\bibinfo {volume} {13}},\
  \bibinfo {pages} {053035} (\bibinfo {year} {2011})}\BibitemShut {NoStop}%
\bibitem [{\citenamefont {Scully}\ and\ \citenamefont
  {Zubairy}(1997)}]{QuantumOpticsScully}%
  \BibitemOpen
  \bibfield  {author} {\bibinfo {author} {\bibfnamefont {M.~O.}\ \bibnamefont
  {Scully}}\ and\ \bibinfo {author} {\bibfnamefont {M.~S.}\ \bibnamefont
  {Zubairy}},\ }\href {https://doi.org/10.1017/CBO9780511813993} {\emph
  {\bibinfo {title} {Quantum Optics}}}\ (\bibinfo  {publisher} {Cambridge
  University Press},\ \bibinfo {year} {1997})\BibitemShut {NoStop}%
\bibitem [{\citenamefont {Tom~Lancaster}(2014)}]{QFT}%
  \BibitemOpen
  \bibfield  {author} {\bibinfo {author} {\bibfnamefont {S.~J.~B.}\
  \bibnamefont {Tom~Lancaster}},\ }\href
  {http://gen.lib.rus.ec/book/index.php?md5=8673f60f9fded3dcb00a475c3fc61999}
  {\emph {\bibinfo {title} {Quantum Field Theory for the Gifted Amateur}}}\
  (\bibinfo  {publisher} {Oxford University Press},\ \bibinfo {year}
  {2014})\BibitemShut {NoStop}%
\bibitem [{\citenamefont {Dicke}(1954)}]{DickeModel}%
  \BibitemOpen
  \bibfield  {author} {\bibinfo {author} {\bibfnamefont {R.~H.}\ \bibnamefont
  {Dicke}},\ }\bibfield  {title} {\bibinfo {title} {Coherence in spontaneous
  radiation processes},\ }\href {https://doi.org/10.1103/PhysRev.93.99}
  {\bibfield  {journal} {\bibinfo  {journal} {Phys. Rev.}\ }\textbf {\bibinfo
  {volume} {93}},\ \bibinfo {pages} {99} (\bibinfo {year} {1954})}\BibitemShut
  {NoStop}%
\end{thebibliography}%
\end{document}